\documentclass{aa}  
\usepackage{graphicx}
\usepackage{txfonts}
\usepackage{comment}
\usepackage{xcolor}
\newcommand{\cn}[1]{\textbf{{\color{violet}{{[CN xxx #1]}}}}}

\newcommand{\E}{\mathcal{E}}
\renewcommand{\P}{\mathcal{P}}
\newcommand{\Pstar}{\P_\star}
\newcommand{\PDM}{\P_{\rm DM}}
\newcommand{\mDM}{m_{\rm DM}}
\newcommand{\Mtot}{M_{\rm tot}}
\newcommand{\mstar}{m_\star}
\newcommand{\Msun}{\,\text{M}_{\odot}}

\newcommand{\rhotot}{\rho_{\rm tot}}
\newcommand{\rt}{r_{\rm t}}

\newcommand{\rGc}{r_{\rm Gc}}

\newcommand{\kms}{{\rm km\,s^{-1}}}

\newcommand{\rp}{r_{\rm peri}}
\newcommand{\rhalf}{r_{\rm half}}
\defcitealias{Pas18}{P18}
\defcitealias{Joh95}{J95}

\begin{document} 
   \title{Tidal mass loss in the Fornax dwarf spheroidal galaxy through N-body simulations with Gaia DR3-based orbits
}
  \titlerunning{Tidal mass loss in the Fornax dSph}
  % \subtitle{}
   \author{Pierfrancesco Di Cintio
          \inst{1,2,3,4}
          \and
          Giuliano Iorio\inst{5,6,7}
          \and
          Francesco Calura\inst{8}
          \and
          Carlo Nipoti\inst{9}
          \and
          Marcello Cantari\inst{9}
          }
   \institute{Consiglio Nazionale delle Ricerche - Istituto dei Sistemi Complessi, Via Madonna del piano 10, I-50019 Sesto Fiorentino, Italy\\
              \email{pierfrancesco.dicintio@cnr.it}
         \and
             INFN-Firenze, via G.\ Sansone 1, I-50019 Sesto Fiorentino, Italy
         \and    
          Dipartimento di Fisica e Astronomia, sezione di Astrofisica e Scienze dello Spazio, Università di Firenze, Piazzale E.\ Fermi 2, I-50125 Firenze, Italy
          \and
          INAF-Firenze, Piazzale E.\ Fermi 5, I-50125 Firenze, Italy
          \and
          Dipartimento di Fisica e Astronomia Galileo Galilei, Università di Padova, Vicolo dell’Osservatorio 3, I–35122 Padova, Italy
          \and
          INFN-Padova, Via Marzolo 8, I–35131 Padova, Italy
          \and
          INAF-Padova, Vicolo dell’Osservatorio 5, I–35122 Padova, Italy
          \and
          INAF-Bologna, via Gobetti 93/3, I-40129, Bologna, Italy
          \and
          Dipartimento di Fisica e Astronomia “Augusto Righi” – Alma Mater Studiorum – Università di Bologna, via Gobetti 93/2, I-40129 Bologna
        }
   %\date{Received September 15, 1996; accepted March 16, 1997}
\date{Draft, August 2, 2024}
% \abstract{}{}{}{}{} 
% 5 {} token are mandatory 
  \abstract
  % context heading (optional)
   {} %leave it empty if necessary  
  % aims heading (mandatory)
   {The Fornax dwarf spheroidal galaxy (dSph) represents a challenge for some globular cluster (GC) formation models, because an exceptionally high fraction of its stellar mass is locked in its GC system. In order to shed light on our understanding of GC formation, we aim to constrain the amount of stellar mass that Fornax has lost via tidal interaction with the Milky Way (MW). }
  % methods heading (mandatory)
   {Exploiting the flexibility of effective multi-component $N$-body simulations and relying on state-of-the-art estimates of Fornax's orbital parameters, we study the evolution of the mass distribution of the Fornax dSph in observationally justified orbits in the gravitational potential of the MW over 12 Gyr.}
   % results heading (mandatory)
   {We find that, though the dark-matter mass loss can be substantial, the fraction of stellar mass lost by Fornax to the MW is always negligible, even in the most eccentric orbit considered. }
   {We conclude that stellar-mass loss due to tidal stripping is not a plausible explanation for the unexpectedly high stellar mass of the GC system of the Fornax dSph and we discuss quantitatively the implications for GC formation scenarios.}
%%%%%%%%%%%%%%%%%%%%%%%%%%%%%%%%%%%%%%%
   \maketitle
%
%-------------------------------------------------------------------
\section{Introduction}
Recent progress in astronomical observations and new instrumentation technology 
have driven a significant improvement in our understanding of composite 
stellar populations in local galaxies. 
In particular, the {\em Gaia} satellite allowed 
us to access some of their most fundamental properties, enabling 
a detailed reconstruction of their phase-space information, possible thanks to the measurement of the {\em Gaia} proper motions of several Milky Way (hereafter, MW) satellites (e.g. see \citealt{2021ApJ...915...48D}).
These new results unveiled the kinematic complexity of various components of our Galaxy and are leading to a detailed characterisation   
of the assembly history of the halo (e.g., \citealt{2019MNRAS.486.3180K,2020MNRAS.495...29E,2021ApJ...919...66B,2023MmSAI..94b.173M,2023A&A...680A..20M}).\\ 
\indent Despite the large wealth of new information, several major questions are still open, 
including the reasons for the present diversity of local dwarf satellites. A few of these pending issues include the mechanisms which 
led to significant differences in their properties, such as,  for example,  their gas content, age of the dominant stellar populations and presence of globular clusters (GCs). 

With regard to this particular point, besides the largest satellites, namely the Large and Small Magellanic clouds, also other, smaller satellites are known to contain old GCs (\citealt{2021MNRAS.500..986H}). In particular, the Fornax dwarf spheroidal galaxy (dSph) is known to host (at least) six GCs (e.g. see \citealt{2019ApJ...875L..13W,2021ApJ...923...77P} and references therein), a number remarkably high with respect to that of other MW satellites.

One peculiar feature of Fornax is its high GC specific frequency  (defined as the number of GCs per unit galaxy luminosity, normalized to a galaxy with an absolute V magnitude of $-15$; e.g.\  \citealt{1999Ap&SS.269..469E}) $S_N\sim 26$\footnote{Local spiral
    galaxies typically have $S_N \le 1$, whereas ellipticals, dwarf ellipticals and
    S0 galaxies show GC specific frequencies in the range $2 \le S_N \le 6$.}, which is exceptionally high for
    satellite galaxies and comparable
    with the ones of cD galaxies at the centres of galaxy clusters, typically showing $S_N>10$.
    Like most GCs, Fornax' GCs  also have multiple stellar populations \citep{2006A&A...453..547L}; moreover,  
    as in MW GCs, even in Fornax GCs there is evidence of 'anomalous' stars that are enriched in helium, nitrogen
and sodium, and depleted in carbon and oxygen, compared to field stars with a more standard abundance pattern \citep{2006A&A...453..547L,2018A&A...613A..56L}. 
This pattern can be produced by invoking p-capture reactions at high temperatures (60-100 million K) occurring in first-generation (FG) stellar polluters.
For such stars, various possibilities have been proposed, including asymptotic giant branch (AGB) stars (\citealt{2008MNRAS.391..825D,2016MNRAS.458.2122D,2019MNRAS.489.3269C}) and
fast rotating massive stars \citep{2007A&A...464.1029D,2008A&A...492..101D}.\\
\indent An issue affecting each of these scenarios is the so-called 'Mass Budget' problem. 
At present the mass fraction of FG stars does not exceed 30-40\% \citep{2015MNRAS.447..927M}. Considering the stellar mass return from a stellar population (e.g, \citealt{2014MNRAS.440.3341C}) and a standard Initial Mass Function (IMF, \citealt{1955ApJ...121..161S,2001MNRAS.322..231K}),
it would be impossible to account for GCs dominated by anomalous stars as the present-day ones from the mass of FG stars only. 
On the other hand, within the AGB scenario, it has been shown that, assuming an initial FG mass that is between 2 and $\sim$ 13 times the one of SG stars,
it is possible to accommodate the FG-to-SG mass observed in GCs \citep{2016MNRAS.458.2122D,2019MNRAS.489.3269C}.\\
\indent In this context, important information was obtained from the study of 
Fornax and its GCs. In particular, \cite{2012A&A...544L..14L} showed that the mass of field stars
can be at most 4-5 times larger than that of its GCs.
If the GCs were originally more massive than today, and if the field stars have all been lost from them,
this result imposes an upper limit on their initial total mass that is lower than the factors required by the AGB scenario. \\
\indent An important quantity that is currently unconstrained is the stellar mass
of Fornax, which in the past could have been greater than today. In principle, 
a fraction of this mass could have been lost as a result of the tidal interaction with the MW.
This possibility was considered by \cite{2015MNRAS.454.2401B} through a series of $N$-body simulations, in which the dynamical evolution  
of Fornax in the gravitational potential of the MW was modelled, considering orbits characterized by different eccentricities.
Their study showed how the tidal effects influence the dark matter (DM) content of the dSph, leading to an amount of DM mass loss up to $\approx40\%$, if measured within a sphere of radius 1.6 kpc. Moreover, in a study based  on 
cosmological simulations, \cite{2016MNRAS.457.4248W} found indications that the peak mass reached by the  DM halo of  Fornax during its evolution was at least 10 times higher than its present-day mass. In this scenario, Fornax must have suffered a significant mass loss, with a peak around 9 Gyr ago, which would coincide with the fall into the Galactic halo.\\
\indent These works, however, where not tailored to study in detail the evolution of the stellar mass
of Fornax, which is of fundamental importance for providing information on the evolution of
its GCs. 
The aim of this work is to study, through $N$-body simulations, the evolution of the stellar and DM distribution of the Fornax dSph orbiting in the MW gravitational potential. 
We will use the method described in  \citet{Nip21} that allows us to represent the dSph with a single component, including both DM and stars, and to study the stellar component 
and the mass loss of the satellite galaxy by means of an a posteriori approach, thus drastically reducing the 
number of simulations to carry out.\\
\indent The paper is structured as follows.  In Sect. \ref{modmeth} we introduce the scheme used to generate multi-component stellar systems from a single species simulation particle, and we present the initial density distribution and the Galactic model assumed for the numerical experiments. We then give some details on the structure of the code used for the simulations and show some numerical tests. In Sect.\ \ref{sec:simu} we present the Fornax-like orbits used in this work and the results of our numerical simulations. In Sect.\ \ref{sec:discussion} we discuss our results and compare them with previous work. Finally, in Sect.\ \ref{sec:conclusions} we summarize and draw our conclusions.
%%%%%%%%%%%%%%%%%%%%%%%%%%%%%%%%%%%%%%
\section{Models and methods}
\label{modmeth}
%%%%%%%%%%%%%%%%%%%%%%%%%%%%%%%%%%%%%%%%%%%%%%%%%%%%%%%%%%%%%%
\subsection{Effective $N$-body models of two-component systems}
Following \citet[][see also \citealt{Bul05,Err15,2021ApJ...909..147S,Bor22}]{Nip21}, in the $N$-body simulations of the dynamics of the Fornax dSph presented in this work we evolve the  satellite as a one-component system (representing the total phase-space distribution), which can be interpreted as an effective two-component system by assigning part of the mass of each particle to the stellar component and the remaining part to the DM component, provided that such mass fractions are conserved throughout the simulation.\\ 
%%%%%%%%%%%%%%%%%%%%%%%%%%%%%%%%%%%%%%%%%%%%%%%%%%%%%%%%%%%%%%%%
\begin{figure}
    \centering
    \includegraphics[width=0.95\columnwidth]{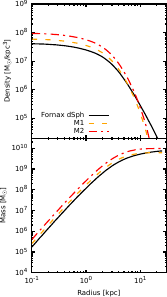}
    \caption{Total density (upper panel) and total mass (lower panel)  profiles of our models of the Fornax dSph progenitor M1 (dashed lines) and M2 (dot-dashed lines),  and of the Fornax dSph (black solid lines) as estimated by \citetalias{Pas18}.}
\label{fig:M1-mass-and-density}
\end{figure}
%%%%%%%%%%%%%%%%%%%%%%%%%%%%%%%%%%%%%%%%%%%%%%%%%%%%%%%%%%%%%%%%%
%%%%%%%%%%%%%%%%%%%%%%%%%%%%%%%%%%%%%%
\begin{table}
\centering
\begin{tabular}{ccc}
\hline
Radius [kpc] &  $\rho_\mathrm{M1}$ [$\mathrm{M}_\odot/{\rm kpc}^3$] & $\rho_\mathrm{M2}$ [$\mathrm{M}_\odot/{\rm kpc}^3$] \\ \hline
0.01 & $6.12\times 10^7$ & $9.71\times 10^7$\\
0.03 & $6.08\times 10^7$ & $9.65\times 10^7$\\
0.1  & $5.91\times 10^7$ & $9.37\times 10^7$\\
0.3  & $5.35\times 10^7$ & $8.48\times 10^7$\\
1.0  & $3.40\times 10^7$ & $5.39\times 10^7$\\
3.0  & $1.03\times 10^7$ & $1.63\times 10^7$\\
10   & $1.03\times 10^5$ & $1.63\times 10^5$\\
25   & $2.56\times 10^1$ & $40.6\times 10^1$ \\ \hline
\end{tabular}
\caption{Total density for the M1 (second column) and M2 (third column) models, as a function of radius (first column).}
\label{tab:density}
\end{table}
%%%%%%%%%%%%%%%%%%%%%%%%%%%%%%%%%%%%%%%
\indent In practice, we first generate a spherical and isotropic $N-$body realization of the stellar system representing the progenitor of Fornax, and we define a stellar portion function $\Pstar(\E)$ which depends on the {\em initial} relative specific energy of the particle $\E=-E>0$ (where $E$ is the particle energy per unit mass). As discussed in \citet{Nip21}, given that $\E$ is an integral of motion for the stationary equilibrium stellar system, in a non-stationary simulation (i.e. the parent model interacting with an external gravitational potential), $\Pstar(\E)$ can be used at any time to assign a stellar mass fraction to each particle, provided that   
$\E$ is computed when the model is set up in equilibrium as an isolated stellar system. The relative energy $\E$ is a tracer of the region in which the particle is located: the higher $\E$ the more likely it is to find the particle in the central regions of the satellite, closer to the centre of its potential well.
Once the stellar mass fraction is defined for a given particle, its associated DM mass fraction is easily obtained as $\PDM(\E)=1-\Pstar(\E)$, so in the simulated $N$-body system a particle of mass $m$ will have stellar mass $\mstar=m\Pstar(\E)$ and  $\mDM=m\PDM(\E)$.\\  
\indent In this framework, all the properties of each of the two components of the $N$-body system, such as for instance the density and velocity distributions, can be computed at any time of the simulation using $\Pstar$ and $\PDM$. The advantage of this approach is that $\Pstar(\E)$, and thus $\PDM(\E$), can be chosen a posteriori. Therefore, a given $N$-body experiment can be reinterpreted in potentially infinite different ways without rerunning the simulation.\\ 
\indent In the present work we adopt as stellar portion function the generalized \cite{1976ApJ...203..297S} function proposed by \citet{Nip21} 
\begin{equation}
\mathcal{P}_{\star}(\E)=\gamma \left( \frac{\E}{\E_{0}}\right)^\alpha \exp{ \left[ - \left( \frac{\E}{\E_{0}}\right)^\beta \right]},
\label{Eq:PiNipoti}
\end{equation}
where $\alpha$, $\beta$ and $\gamma$ are dimensionless parameters, and $\E_{0}$ is a scale specific energy. As discussed by \citet{Nip21}, where the individual effects of varying, $\alpha$, $\beta$, $\gamma$ and $\E_0$ are illustrated, 
the portion function (\ref{Eq:PiNipoti}) turns out to be quite flexible, and allows one to produce systems with a wide variety of relative stellar and DM density distributions, for a given total density distribution.
%%%%%%%%%%%%%%%%%%%%%%%%%%%%%%%%%%%%%%%%%%%%%%%%%%%
\subsection{$N$-body models of Fornax's progenitor}
\label{sec:nbody_models}

In this work we present simulations of a satellite galaxy, representing the Fornax dSph, in orbit around the MW from an initial time 12 Gyr in the past up to the present day. 
 For this purpose, we require that the total final density profile of the satellite is consistent with the present-day estimated total mass distribution of Fornax. During the dynamical evolution in the MW potential, the satellite is expected to lose mass via tidal stripping. It is therefore natural to consider as initial condition a stellar system that is more massive than the present-day Fornax dSph, but with a total density profile with a similar shape in the central parts.\\   
\indent We assume as reference present-day total density distribution of Fornax the best-fit model obtained with dynamical modelling by \citet[][hereafter \citetalias{Pas18}]{Pas18}. Such model is a spherical two-component system constructed from analytic distribution functions, whose stellar and DM profiles are then evaluated numerically. The total density profile $\rho_{\rm tot,P18}(r)$ is the sum of the best-fitting stellar and DM density profiles shown in figure 3 of \citetalias{Pas18}.\\ 
\indent The  total density profile of  Fornax's progenitor is obtained by modifying the density profile $\rho_{\rm tot,P18}$ as
\begin{equation}\label{rhoincond}
\rhotot(r)=A\rho_{\rm tot,P18}(r)\exp{\left[-\left(\frac{r}{\rt}\right)^2\right]}, 
\end{equation}
where $A$ is a dimensionless scale factor and $\rt$ is a truncation radius. In particular, we consider two models for the progenitor:
model M1, with $A=1.45$ and $\rt=10$ kpc (and thus total mass $\Mtot=7.18\times 10^{9}\Msun$), and model M2, with $A=2.3$ and $\rt=10$ kpc (and thus total mass $\Mtot=1.13\times 10^{10}\Msun$). Both models have half-mass radius $\rhalf\simeq 4.3$ kpc. In Fig. \ref{fig:M1-mass-and-density} we show the total density and mass profiles of M1 and M2, together with the best-fitting present-day Fornax total density and mass profiles of \citetalias{Pas18}, while in Tab. \ref{tab:density} we tabulate the density profiles of models M1 and M2. 

In the simulations we use as initial conditions isotropic $N$-body realizations of the models M1 and M2 produced using the Python module {\sc{OpOpGadget}}\footnote{\url{https://github.com/giulianoiorio/OpOpGadget}} developed by G.\ Iorio.
The total density profile $\rho_{\rm tot,P18}$ is formally untruncated, but it must be stressed that  beyond $\approx 3$ kpc the stellar mass of Fornax is negligible (e.g.\ \citetalias{Pas18}), and the DM distribution is poorly constrained, because there is no luminous tracer to probe the mass distribution at $r\gtrsim 3$ kpc. Thus, provided that the truncation radius $\rt$ is sufficiently larger than $\approx 3$ kpc, the evolution of the stellar component of the satellite is expected to be independent of the specific value of $\rt$ adopted in the $N$-body realization. In practice, the exponential cut-off (Eq.~\ref{rhoincond}) with $\rt=10$ kpc is applied to avoid to sample with particles regions very far from the bulk of the stellar distribution, which cannot be probed observationally. As far as the DM mass loss is concerned, we will consistently consider only the mass within 3 kpc from the satellite's centre, whose evolution is essentially unaffected by the specific choice of $\rt$. 

\subsection{The Milky Way gravitational potential model}
\label{sec:mwpot}

In our simulations we model the MW as a frozen gravitational potential, adopting in particular the \citet[][\citetalias{Joh95}]{Joh95} axisymmetric model. Compared to other MW models (e.g.\ \citealt{2005MNRAS.364..433B,2014MNRAS.445.3788G,2014MNRAS.445.3133P,2019MNRAS.487.4025C} and references therein), the \citetalias{Joh95} model can be considered as a ``heavy" MW, having  mass of $\simeq 5  \times 10^{11} \Msun$ within 50 kpc and $\simeq 2.3 \times 10^{12} \text{ M}_{\odot}$ within 300 kpc \citep[see][]{Ior19}. Thus, adopting the \citetalias{Joh95} potential, in our simulations we tend to maximize the effects of tidal stripping on Fornax.\\ 
\indent The \citetalias{Joh95} MW model consists of three components (disk, bulge and halo) with total gravitational potential $\Phi_{\rm MW,tot}(R,z)=\Phi_{\text{disk}}+\Phi_{\text{bulge}}+\Phi_{\text{halo}}$, where
\begin{equation}
   \Phi_{\text{disk}}(R,z)= -\frac{GM_{\text{disk}}}{\sqrt{R^2 + \big(a+\sqrt{z^2 + b^2}\big)^{2}}}
   \label{eq:MijaNagai} 
\end{equation}
for a \citet{MiyamotoNagai} disk,
\begin{equation}
    \Phi_{\text{bulge}}(\rGc)= -\frac{GM_{\text{bulge}}}{\rGc+c}
    \label{eq:JohnHern}
\end{equation}
for a spherical \citet{Her90} bulge,
and 
\begin{align}
     \Phi_{\text{halo}}(\rGc)= v^2_{\text{halo}} \ln{( \rGc^2 +d^2 )}
     \label{eq:JohnHalo}
\end{align}
for  a spherical \cite{1981MNRAS.196..455B} logarithmic halo.
In Equations (\ref{eq:MijaNagai}), (\ref{eq:JohnHern}) and (\ref{eq:JohnHalo}), $R=\sqrt{x^2+y^2}$ and $z$ are Galactocentric cylindrical coordinates, and $\rGc=\sqrt{R^2+z^2}$ is the Galactocentric distance, while the scale and mass parameters are $a=6.5$ kpc, $b=0.26$ kpc, $c=0.7$ kpc, $d=12$ kpc, $M_{\text{bulge}}=3.4\times 10^{10}\text{ M}_{\odot}$, $M_{\text{disk}}=10^{11}\text{ M}_{\odot}$ and $v_{\text{halo}}=128$ km/s.\\
\indent  
At variance with some recent works (e.g.\ \citealt{2021MNRAS.501.2279V,GAIADR3,2023MNRAS.518..774L,2023MNRAS.521.4936K}), we do not account for the contribution of the Large Magellanic Cloud (LMC)  to the Galactic potential (see also \citealt{Garavito-Camargo_2021}). However, as shown in figure D1 of \cite{GAIADR3}, the perturbations induced by the LMC can affect the orbit of Fornax for different initial conditions, but on average its pericentric radius $r_{\rm peri}$ is only slightly altered. Given that  the mass loss via tidal stripping is expected to be maximal around $r_{\rm peri}$, we can safely assume that neglecting the LMC in a heavier MW model does not influence significantly the mass loss of the Fornax dSph with respect to more realistic models including the effect of the LMC (\citealt{2024MNRAS.527..437V} and references therein).
%%%%%%%%%%%%%%%%%%%%%%%%%%%%%%%%%%%%%%%%%%%%%%%%%%%
\begin{figure}
        \centering
        \includegraphics[width=\columnwidth]{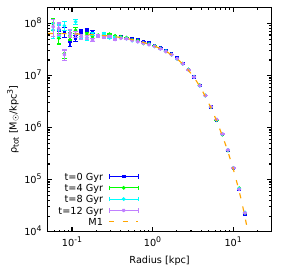}
    \caption{Total density profiles of representative snapshots of the $N$-body simulation in which an $N$-body realization of model M1 is evolved in isolation. The dashed line marks the reference initial profile (Eq.\ \ref{rhoincond} with the parameters of model M1). The error bars are computed considering the Poissonian uncertainty on particle counts.}
\label{Fig:pinipoti_is}
\end{figure}
%%%%%%%%%%%%%%%%%%%%%%%%%%%%%%%%%%%%%%%%%%%%%%%%%%
 \subsection{$N$-body code and numerical tests}
\label{testsrho}

The simulations presented in this work have been performed with the collisionless $N$-body code {\sc fvfps} ({\sc fortran} version of a fast Poisson
solver, \citealt{2003MSAIS...1...18L,2003MNRAS.342..501N}), based on \cite{2002JCoPh.179...27D} Poisson solver.
As a rule, we adopt a softening length $\epsilon=3.1\times 10^{-2}$ kpc. We set the minimum value of the opening parameter $\theta_{\rm min}$, used to
control the accuracy of the force evaluation within the tree scheme, to 0.5.\\
\indent The contribution of the MW is implemented adding the gravitational field derived from the potential $\Phi_{\rm MW,tot}$ (Sect.\ \ref{sec:mwpot}) to the self-consistent $N-$body gravitational field $-\nabla\Phi_{\rm self}$.
The equations of motion 
\begin{equation}\label{eom}
\ddot{\mathbf{r}}_i=-\nabla[\Phi_{\rm self}(\mathbf{r}_i)+\Phi_{\rm MW,tot}(\mathbf{r}_i)]
\end{equation}
of a particle with position vector $\mathbf{r}_i$
are propagated with a second order leapfrog scheme using a fixed time-step that we set to $0.01t_{\rm dyn}$, where 
\begin{equation}
t_{\rm dyn}\equiv \sqrt{\frac{8\pi\rhalf^3}{3GM_{\rm tot}}}
\end{equation}
 is the half-mass dynamical time of the satellite in the initial conditions. 

%Based on some numerical test, we present results of simulations with number of particles $N=10^6$.

As preliminary tests, we ran simulations of the $N$-body realizations of models M1 and M2 evolved in isolation for 12 Gyr. For both models we used different resolutions with $N$ spanning a range between $10^4$ and $3\times 10^6$. We found that the isolated $N$-body systems maintain very well their equilibrium, when simulated with $N\gtrsim 10^5$. The simulations discussed in this work have been performed with $N=10^6$, that gives a good balance between computational time and mass resolution.
As an example, in Fig.\ \ref{Fig:pinipoti_is} we show the total density profiles of the initial conditions and of three representative snapshots of the simulation of the isolated M1 model with $N=10^6$. The density profile of the simulated stellar system remains close to the initial profile for the entire time span  of the simulation. Small fluctuations below $r\lesssim 10^{-1}$ kpc $\approx 3\epsilon$ are a consequence of shot noise due the poorer sampling in the inner regions, as shown also by the wider Poissonian error bars. Based on these tests, we can robustly ascribe to tidal effects any significant variation of the mass distribution of the $N$-body systems evolved in the presence of the MW potential.
 %%%%%%%%%%%%%%%%%%%%%%%%%%%%%%%%%%%%%%%%%%%%
\subsection{Limitations of the models}

Our models of the evolution of the Fornax dSph in the MW are simplified
in some aspects, which we summarize and discuss here.

\begin{itemize}

\item As mentioned in Section~\ref{sec:mwpot}, it is likely that the adopted MW model is unrealistically heavy. Fornax's orbit and its tidal stripping history would be different in more
realistic, lighter MW models. The gravitational potential of the MW is also assumed to remain the same over the last 12 Gyr, while in the standard hierarchical scenario of structure formation we expect that its mass gradually grows over time via accretion and merging. Moreover, as in our simulations the MW is not represented with particles, the dynamical friction it exerts on the satellite is neglected, so the apocentric distances of the simulated Fornax are underestimated (see Appendix \ref{sec:app_dyn_fric}). All the aforementioned features of the MW model are expected to have the effect of increasing the tidal stripping, because in a more realistic situation (lighter MW and/or operative dynamical friction) the satellite would tend to orbit in regions with weaker tidal forces. 

\item Fornax is modelled as a collisionless system for 12 Gyrs, even if this dwarf galaxy is known to have had a prolonged star formation history that ceased only less than 1 Gyr ago (e.g.\ \citealt{deB12}). It follows that Fornax must have had, over most of its lifetime, a gaseous component, which is not included in our models. Though gas behaves differently from the collisionless components during the satellite's orbit around the MW (for instance, because it suffers also ram-pressure stripping), its presence should not affect significantly our results, because we are interested in the evolution of the {\em stellar} mass and not of all the {\em baryonic} mass. 
It is also the case that, in our approach, we do not follow the build-up of the stellar component of Fornax over cosmic time, but, de facto, we assume that all the stars are in place since 12 Gyr ago. Doing so,  we tend to enhance the chance that the stars are tidally stripped, because they are exposed to the tidal field of the MW for longer time.

%especially because there are indications that in Fornax the younger stars have a more concentrated spatial distribution \citep{Bat06}.

\item

A limitation of the energy-based effective two-component $N$-body model is that the stellar and DM component share the same isotropic velocity distribution. The effectiveness of tidal stripping is known to depend on the orbits of the stars within the satellite \citep{Kaz04,Rea06}, so the fractions of stellar and DM mass lost by Fornax could be different if the two components were allowed to have different velocity distributions. However, at least for the stellar component, the assumption of isotropic distribution function should not be too restrictive, because in the present-day stellar component of Fornax there is no evidence that the velocity distribution deviates from isotropy (\citetalias{Pas18}, \citealt{Kow19}).

\end{itemize}

The aforementioned simplifying assumptions are justified
because the main aim of our study is addressing the specific question of
how much stellar mass, at most, Fornax can have lost ceding stars to
the MW. Our results are thus robust if they are used to address this question,
but cannot provide information on other aspects of the evolution
of the Fornax dSph, such as, for instance the star formation history or even the precise stellar mass content at given look-back time. 
In the following we will thus focus primarily on the question of the fraction of stellar mass loss, for which we can robustly estimate an upper limit. We take advantage of our simulations also to gain insight on some aspects of the evolution of the DM halo of Fornax. In Section~\ref{sec:massloss} we  quantify the fraction of DM mass lost, but limiting ourselves only to the DM contained within the region in which the dynamical mass can be probed observationally. In Section~\ref{sec:rho150} we consider the evolution of the central DM density of the satellite:  however, in this respect, our purpose is limited to studying the effect of the pericentric passages on the DM density distribution, with no attempt to include the complex effects of baryon physics on the inner DM distribution in dwarf galaxies \citep[e.g.][and references therein]{Nip15,Mun24}.
\section{Dynamical evolution of Fornax orbiting the Milky Way}
\label{sec:simu}
%%%%%%%%%%%%%%%%%%%%%%%%%%%%%%%%%%%%%%%%%%%%%%%
\begin{figure}
\centering
\includegraphics[width=\columnwidth]{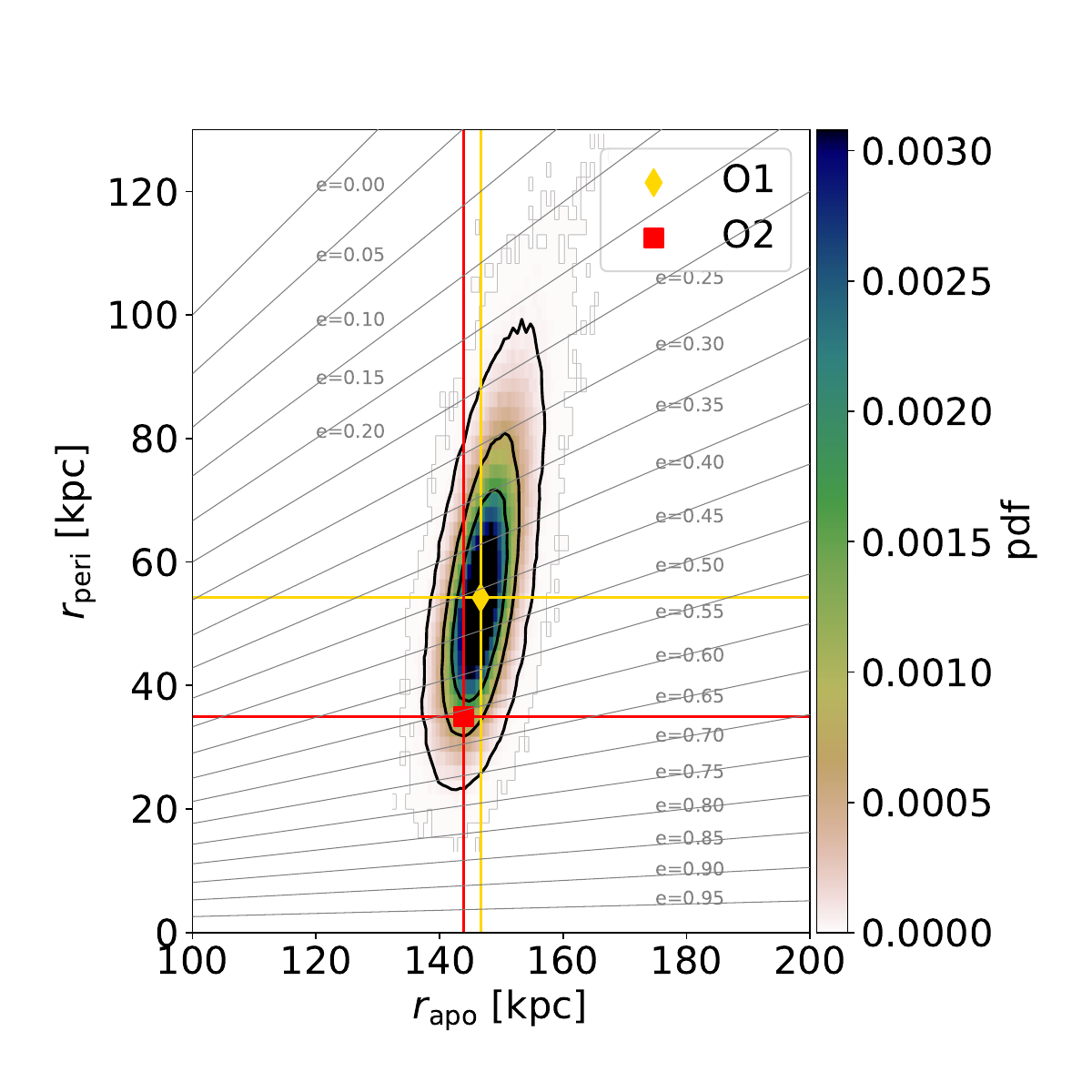}
\caption{Distribution of Fornax apocentric and pericentric radii from the orbit integration (colour map). The contour levels (thick solid lines) show the area containing the 39\%, 86\%  and 99\%  of the distribution (corresponding to 1-$\sigma$, 2-$\sigma$ and 3-$\sigma$ of a 2D Gaussian). The oblique grey lines indicate orbits of a given eccentricity ranging from $e=0$ to $e=0.95$ with a step $\delta e=0.05$.}
\label{fig_incond}
\end{figure}
%%%%%%%%%%%%%%%%%%%%%%%%%%%%%%%%%%%%%%%%%%%%%%%%%%%
%%%%%%%%%%%%%%%%%%%%%%%%%%%%%%%%%%%%%%%%%%%%%%%%%%
\begin{figure*}
\centering
\includegraphics[width=0.95\textwidth]{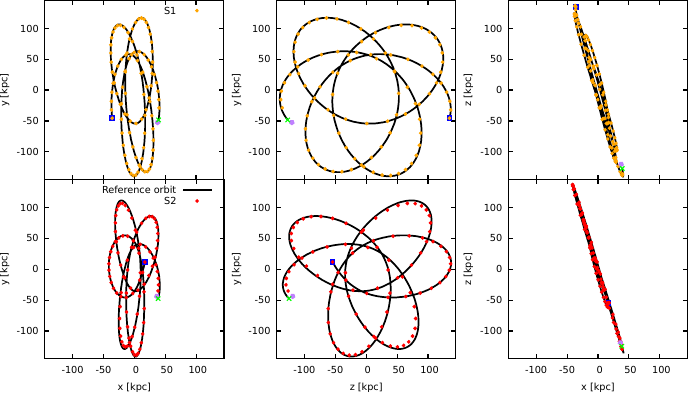}
\caption{$x-y$, $z-y$ and $x-z$ projections of Fornax-like O1 (upper panels) and O2 (lower panels) orbits of a point-like mass in the J95 potential (solid lines), and of the trajectories of the satellite's centre in the corresponding S1 and S2 simulations (yellow and red dots, respectively). The phase-space coordinates of the initial conditions (blue squares) are given in Tab.~\ref{tab:init}. The green crosses indicate the observationally determined present-day position of Fornax, which coincides, by construction, with the end point of the reference point-mass orbits. The purple pentagons mark the present-day position of the centre of the satellite in the $N$-body simulations.}
\label{figprojections}
\end{figure*}
%%%%%%%%%%%%%%%%%%%%%%%%%%%%%%%%%%%%%%%%%%%%%%%%%%%%%%%%%%%%%%
%%%%%%%%%%%%%%%%%%%%%%%%%%%%%%%%%%%%%%%%%%%%%
\begin{figure*}
    \centering
    \includegraphics[width=0.85\textwidth]{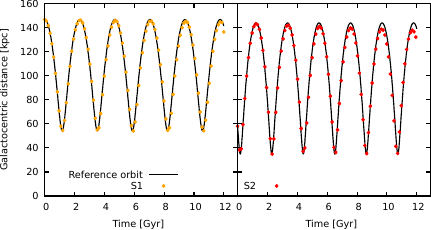}
    \caption{Galactocentric distance of the satellite's centre as function of time for simulations S1 (left) and S2 (right), respectively. The solid lines mark the trajectory of a point-mass particle with the same initial conditions as the c.o.m.\ of the satellite in the two simulations.}
    \label{fig:S1-CmVStime}
\end{figure*}
%%%%%%%%%%%%%%%%%%%%%%%%%%%%%%%%%%%%%%%%%%%%
%%%%%%%%%%%%%%%%%%%%%%%%%%%%%%%%%%%%%%%%%%%
\begin{figure*}
\centering
\includegraphics[width=0.75\textwidth]{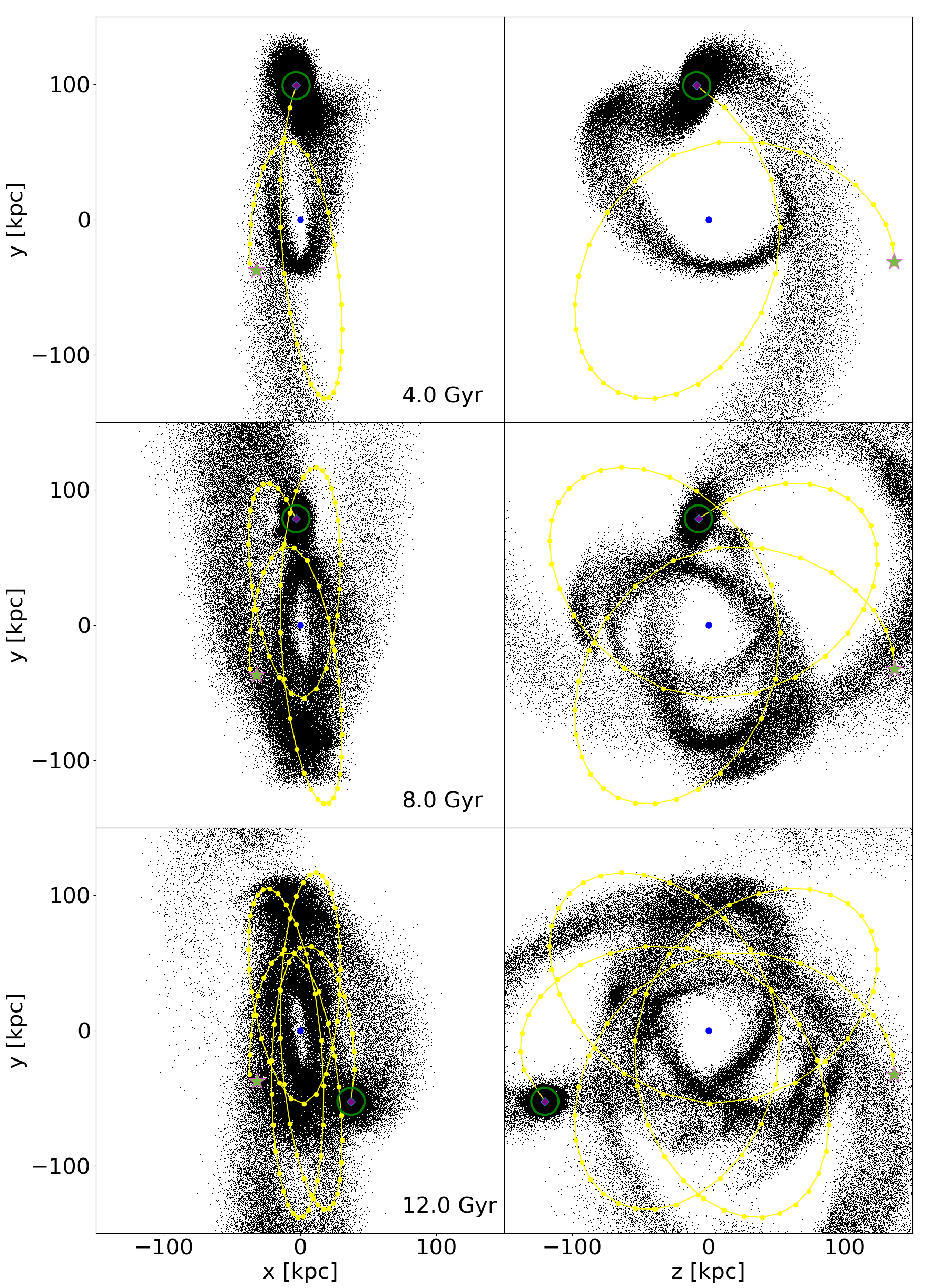}
\caption{Spatial distribution of the particles of the satellite in the $x-y$ and $z-y$ planes at $t=4$, 8 and 12 Gyr for the S1 simulation. Green star: initial position of the satellite centre. Dotted lines: trajectory of the of satellite's centre. Blue dot: centre of the MW model. Green circle: 10 kpc radius circumference centred in the satellite's centre at the time of the snapshot. Here the particles represent the total  (stars plus DM) mass distribution.}
\label{fig:M1-Pannel}
\end{figure*}
%%%%%%%%%%%%%%%%%%%%%%%%%%%%%%%%%%%%%%%%%%%%%%%%%%%% 
%%%%%%%%%%%%%%%%%%%%%%%%%%%%%%%%%%%%%%%%%%%
\begin{figure*}
\centering
\includegraphics[width=0.75\textwidth]{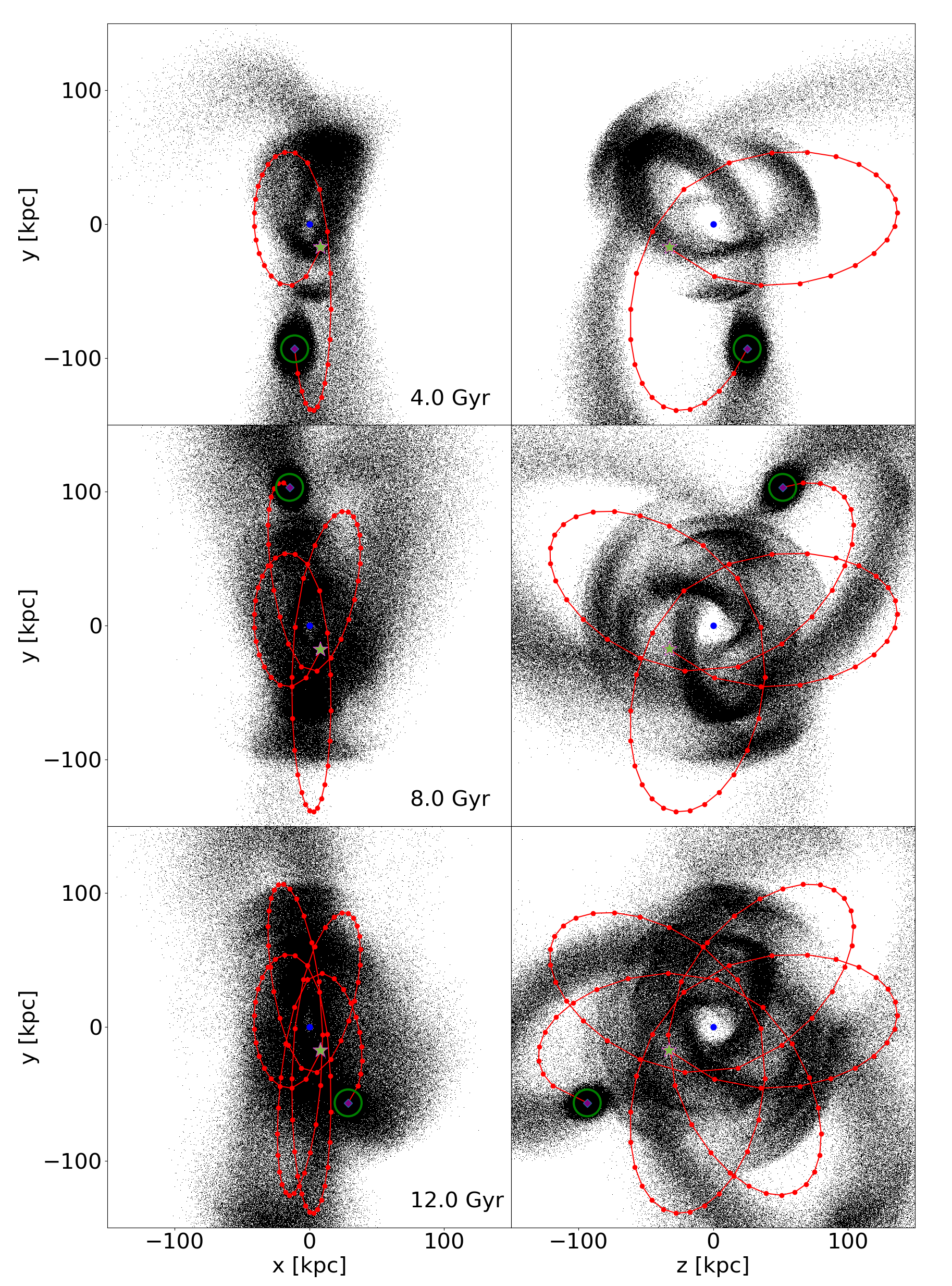}
\caption{Same as in Fig. \ref{fig:M1-Pannel}, but for the simulation S2.}
\label{fig:M2-Pannel}
\end{figure*}
%%%%%%%%%%%%%%%%%%%%%%%%%%%%%%%%%%%%%%%%%%%%%%%%%%%% 
%%%%%%%%%%%%%%%%%%%%%%%%%%%%%%%%%%%%%%%%%%%%%%%%%%%%%%%%%%%
\begin{table*}
\centering
\begin{tabular}{llllllllll}
\hline
Orbit name        & $x$ [kpc]      & $y$ [kpc]      & $z$ [kpc]   & $v_x$ [$\kms$]      & $v_y$ [$\kms$]      & $v_z$ [$\kms$]    & $\rp$ [kpc] & $r_{\rm apo}$ [kpc] & $e$    \\ \hline
O1 (initial)    & -35.76  & -45.49   & 134.76  &  -12.90 &  104.11 & 31.4   & 54.2 & 146.6 & 0.467      \\ 
O2 (initial)    &  17.09  & 11.60    & -54.23  &  -52.69 & -238.52 & 127.4  & 34.9 & 143.8 & 0.613   \\ \hline
O1 (present day)    & 39.17  & -48.29   & -126.94  &  -14.53 &  -92.15 & 77.29   &  &  &       \\ 
O2 (present day)    &  38.46 & -47.50    & -124.85  &  -19.76 & -65.69 & 68.74  &  &  &    \\ \hline
\hline
\end{tabular}
\caption{Initial conditions (top rows) and present-day phase space coordinates of the O1 and O2 orbits. $x$, $y$, $z$ are the Galactocentric Cartesian coordinates and $v_x$, $v_y$, $v_z$ are the corresponding velocities. $r_{\rm peri}$, $r_{\rm apo}$ and $e$ are the pericentric and apocentric distances and orbital eccentricity, respectively.}
\label{tab:init}
\end{table*}
%%%%%%%%%%%%%%%%%%%%%%%%%%%%%%%%%%%%%%%%%%%%%%%%%%%%%%%%%%%%%%%%%%%%%%%%%%%%%%%%%%%%%%%%%%%%%%%%
\begin{figure*}
\centering
\includegraphics[width=\textwidth]{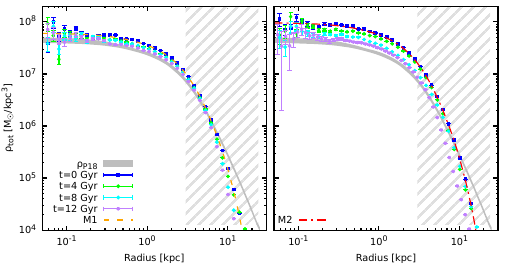}
\caption{Evolution of the total density profile over 12 Gyr in simulations S1 (left) and S2 (right). The symbols have the same meaning as in Fig. \ref{rhoincond}. The red and yellow dashed lines mark the initial profiles (models M1 and M2). The gray shaded area is the present-day total density profile of the Fornax dSph estimated by \citet{Pas18} with its 1-$\sigma$ uncertainty. The hatched parts of the graphs indicate regions where  the total mass distribution of the Fornax dSph is poorly constrained, because there are no luminous tracers of the gravitational potential.}
\label{figevolrho}
\end{figure*}
%%%%%%%%%%%%%%%%%%%%%%%%%%%%%%%%%%%%%%%%%%%%%%%%%%%%%%
\subsection{Orbital parameters and $N$-body simulations}
To study the dynamical evolution of the Fornax-like satellite in the MW, we consider two initial conditions for the satellite's centre of mass, corresponding to two different orbits in the \citetalias{Joh95} MW gravitational potential that we will refer to as orbits O1 and O2.
 In order to define these two orbits, we take the Fornax Gaia-DR3 proper motions, heliocentric distance and sky position from \citet[][]{GAIADR3}, $\mu_\alpha=0.381 \pm 0.001$ mas yr$^{-1}$, $\mu_\delta=-0.358 \pm 0.002$ mas yr$^{-1}$,  $D_\odot=139.3 \pm 2.6$ kpc, $\alpha \mathrm{(RA)=39.96667^\circ}$ and $\delta \mathrm{(DEC)=-34.51361^\circ}$ (assumed without errors), and the systemic line-of-sight velocity from \citet[][]{Bat06}, $v_\mathrm{sys}=54.1 \pm 0.4$ km s$^{-1}$. We retrieve the intrinsic position and velocity of Fornax in the Galaxy, defining the galactic frame of reference as described in \cite{Ior19}, by assuming that the Sun is at a distance $R_\odot=8.13 \pm 0.3$ kpc from the Galactic centre \citep{gravity2018}, the velocity of the local standard of rest (lsr) is $v_\mathrm{lsr}=238 \pm 9$ km s$^{-1}$ \citep{Schonrich2012} and the Sun peculiar motion with respect to the lsr is $(U_\odot,V_\odot,W_\odot)=(-11.1 \pm 1.3, 12.2 \pm 2.1, 7.25 \pm 0.6)$ km s$^{-1}$ \citep{Schonrich2010}. To estimate the effect of the parameter uncertainties on the orbit estimate, we sampled 10 millions orbits randomly drawing the values of the aforementioned parameters, assuming that the associated errors are Gaussian and adding a systematic proper-motion error of 0.017 mas yr$^{-1}$ \citep{GAIADR3}. In Fig.\ \ref{fig_incond} we show the resulting distribution of pericentric ($r_{\text{peri}}$) and apocentric ($r_{\text{apo}}$) radii. The orbit O1, which can be considered the "fiducial" orbit of Fornax, has $\rp$ and $r_{\text{apo}}$ corresponding to the peak of the distribution shown in Fig.\ \ref{fig_incond}.  
As representative of an extreme orbit, the O2 is chosen among those with the smallest pericentres considering the orbits within 2-$\sigma$ of the fiducial value in the joint distribution of $r_{\text{peri}}$ and $r_{\text{apo}}$ (see Fig. \ref{fig_incond}). Both O1 and O2 are rather eccentric with, respectively, $e\simeq 0.5$ and $e\simeq0.6$, where  the 
eccentricity $e$ is defined in the standard way as
\begin{equation}
e = \frac{r_{\text{apo}} - r_{\text{peri}}}{r_{\text{apo}} + r_{\text{peri}}}.
\end{equation}
The initial conditions for the orbits O1 and O2, summarized in Tab. \ref{tab:init} are recovered by integrating backwards in time using the {\sc Python} package {\sc galpy} \citep{Bov15} for 12 Gyr in the \citetalias{Joh95} gravitational potential, starting from the corresponding sets of present-day phase-space coordinates, also given in Tab. \ref{tab:init}.\\
\indent We present here the results of two simulations, dubbed S1 and S2, in which we have followed for 12 Gyr the evolution of $N$-body realizations of the model M1 and M2, starting from the initial conditions of orbit O1 and O2, respectively. To compare the simulated trajectory of the Fornax dSph progenitor with the corresponding reference orbits, in each run we have evaluated as a function of time the position of the density peak\footnote{As it is usual in simulations with substantial tidal stripping, the centre of mass cannot be used as centre of the satellite, because of its dependence on the geometry of the tidal tails.}
of the $N$-body system, which, hereafter, we will refer to as the centre of the satellite. The latter is determined following the iterative method of \cite{Pow03}.
% with an iterative technique that evaluates the centre of mass of particles within a sphere that shrinks recursively until the convergence. At each iteration, the centre of the sphere is reset in centre of mass of the previous step and its radius is reduced by the 2.5\%. The procedure halts once a specified fraction of particles (typically the 1\%) is reached within said sphere
In Fig.\ \ref{figprojections} we plot the trajectories of the $N$-body satellite's centre in simulations S1 and S2, against the corresponding point-mass orbits O1 and O2 in the $x-y$, $z-y$ and $x-z$ planes.
The orbit of the Fornax progenitor in the simulation S1 appears to follow closely the corresponding point-mass trajectory O1 for the whole 12 Gyr-integration. The present-day position of the centre of the density distribution (marked in the figure with the purple pentagon) falls very close to its parent position in O1 (indicated by the green cross). The orbit of the (initially) heavier model M2 in simulation S2 starts to show significant discrepancies with respect to its reference orbit O2 at later stages of the evolution (say at around $\sim 7$ Gyr) due to the so-called dynamical self-friction\footnote{The effect of the dynamical friction exerted by the host galaxy on the satellite is  absent by construction in our simulations in which the host system is modelled as a fixed gravitational potential {(see Appendix~\ref{sec:app_dyn_fric} for a discussion)}.} (\citealt{Mil20}) exerted by the tidal tails. 
The Galactocentric distance (i.e.\ the distance from the Galactic centre of the satellite's centre) as a function of time is shown in Fig.\ \ref{fig:S1-CmVStime} for simulations S1 and S2 . While in simulation S1 (left panel) no appreciable decay of the apocentre is observable, in simulation S2 (right panel) the apocentre of the orbit has decayed of about 5 kpc after $7$ Gyr.
%%%%%%%%%%%%%%%%%%%%%%%%%%%%%%%%%%%%%%%%%%%%%%%%%%%%%%%
\begin{figure*}
\includegraphics[width=0.9\textwidth]{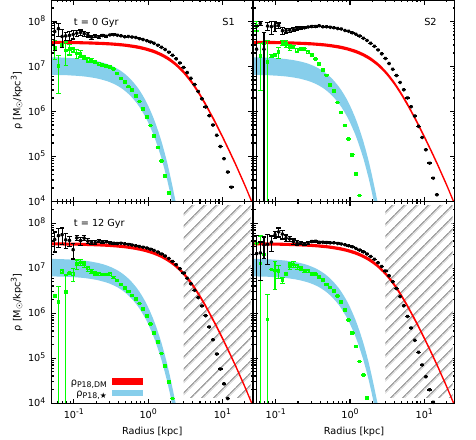}
\caption{DM (black diamonds) and stellar (green squares) density profiles of the S1 (left) and S2 (right) simulations at $t=0$ (upper panels) and 12 Gyr (lower panels). In all panels the cyan and red shaded areas represent, respectively, the stellar and DM density profiles of the Fornax dSph within 1-$\sigma$ uncertainty, as estimated by \citetalias{Pas18}. The hatched region has the same meaning as in Fig.\ \ref{figevolrho}}
\label{fig:S1-Density-Prof}
\end{figure*}
%%%%%%%%%%%%%%%%%%%%%%%%%%%%%%%%%%%%%%%%%%%%%%%%%%%%%%%   
\subsection{Evolution of the satellite total mass distribution}
\label{sec:totalmassevo}
In Figs.\ \ref{fig:M1-Pannel} and \ref{fig:M2-Pannel} we show snapshots of the particle distributions in the $x-y$ and $z-y$ planes at different times ($t=4$, 8 and 12 Gyr) in the simulations S1 and S2, respectively, as well as the trajectories up to said times, as indicated by the yellow and red dotted lines. In each panel, to guide the eye, we superimpose a circle of radius 10 kpc centred in the satellite's centre at the time of the snapshot. In both simulations, already through the first 4 Gyr of integration the $N-$body systems have developed long tidal tails. We recall that the particles shown in these figures trace the total mass distribution, with no distinction between stars and DM.\\
\indent In Fig. \ref{figevolrho} we show the evolution of the total density profiles of the satellite in simulations S1 and S2. As a reference, we added in both panels the initial distribution of the parent smooth model (Eq.\ \ref{rhoincond}) and its $N-$body realization, as well as  the profile estimated by \citetalias{Pas18} as the present-day total density distribution of the Fornax dSph with its 1-$\sigma$ uncertainty.\\
\indent The  final angle-averaged\footnote{The density profiles have been evaluated within spherical shells, which is justified because the final axial ratios within a sphere enclosing the 80\% of the total initial mass (see \citealt{2006MNRAS.370..681N} for the numerical procedure) are $c/a=b/a=0.99$ for S1 and $c/a=0.99$ and $b/a=0.98$ for S2, where $c\leq b\leq a$ are the principal axes of the inertia tensor.} total density profiles of the simulation end-products at 12 Gyr appear compatible with $\rho_{P18}$ below 3 kpc, which is the radius within which we have luminous kinematic tracers. The values of the density in simulation S1 are within  1-$\sigma$, while in simulation S2 are slightly above 1-$\sigma$ of the estimate of \citetalias{Pas18}.
%%%%%%%%%%%%%%%%%%%%%%%%%%%%%%%%%%%%%%%
\begin{figure*}
        \centering
        \includegraphics[width=0.85\textwidth]{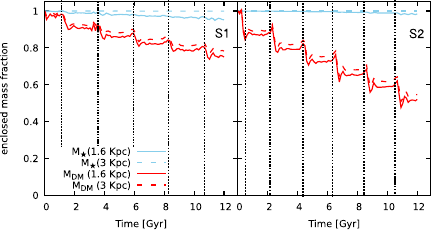}
\caption{Evolution of the stellar (red lines) and DM (cyan lines) mass fractions of the satellite enclosed within spheres of radius of 1.6 (solid lines) and 3 kpc (dashed lines) for the simulations S1 (left) and S2 (right). 
The vertical dashed lines mark the times of the passages at the pericentre.}
\label{fig:S1-MassLoss}
\end{figure*}
%%%%%%%%%%%%%%%%%%%%%%%%%%%%%%%%%%%%%%%%%%%%%%%%%%%%%%%%%%%%%%%%

\subsection{Evolution of the satellite stellar and dark-matter density distributions and mass loss}
\label{sec:massloss}

As discussed in Section \ref{modmeth}, we inferred the final stellar and DM distributions of the simulations by tuning the values of the parameters of the portion function (\ref{Eq:PiNipoti}), so that the final stellar density distribution matches as far as possible the present-day stellar mass distribution of Fornax. In particular, we take as reference the 3D stellar density profile of the best-fitting \citetalias{Pas18} model.\\ 
\indent 
For each simulation, we found empirically a combination of values of the parameters
$\alpha$, $\beta$, $\gamma$ and $\mathcal{E}_0$ of the portion function (\ref{Eq:PiNipoti}) such that the final stellar density profile of the simulated satellite is consistent with that estimated for Fornax . The values of the parameters are  given in Tab. \ref{tab:SIM_PROP}.
%%%%%%%%%%%%%%%%%%%%%%%%%%%%%%%%%%%%%%
\begin{table}
\begin{tabular}{lllll}
\hline
Simulation &  $\alpha$ & $\beta$ & $\gamma$ & ${\mathcal{E}}_0$  [km$^2$/s$^2$] \\ \hline
S1 & 30 & 1.0 & 1.52  &  $ 8.52\times 10^3 $\\ \hline
S2  & 65 & 0.1 & 0.02 &  $ 1.37\times 10^4 $\\ \hline
\end{tabular}
\caption{Parameters of the portion function (\ref{Eq:PiNipoti}) adopted a posteriori for the two simulations presented in this work.}
\label{tab:SIM_PROP}
\end{table}
%%%%%%%%%%%%%%%%%%%%%%%%%%%%%%%%%%%%%%%
The resulting final stellar and DM density profiles are shown in Fig.\ \ref{fig:S1-Density-Prof} (lower panels). For the simulation S1, the final stellar density profile matches rather well that of \citetalias{Pas18} within its error bars (blue shaded area) out to $\sim 2.3$ kpc. In S2, the stellar density profile falls within the error bars of the reference density only out to $\sim 1$ kpc, while at larger radii the final simulated system has stellar density profiles truncated slightly more sharply than in \citetalias{Pas18} model (lower right panel). 
As noted for the total density (see Sect.\ \ref{sec:totalmassevo}), in both cases, the DM profile is slightly higher than  that of the reference model (red shaded area) out to about 4 kpc, to fall considerably below it at larger radii. We recall, however, that the comparison with P18 is meaningful only out to $\approx 3$ kpc. 
In the upper panels of Fig. \ref{fig:S1-Density-Prof} we also show the initial density profiles of stars and DM. We observe that, in particular for S1, the initial stellar density is characterized by a steeper cusp than its counterpart after 12 Gyr of evolution. As it happens for to the total density profiles, the inner profile, below $10^{-1}$ kpc ($\approx 3\epsilon$) is more noisy and has wider error bars\footnote{The errors on the stellar and DM density are computed, respectively, as 
$\Delta\rho_{\star}=(\rho_\star/\rho_{\rm tot})\Delta\rho_{\rm tot}$ and $\Delta\rho_{\rm DM}=(\rho_{\rm DM}/\rho_{\rm tot})\Delta\rho_{\rm tot}$, where $\Delta\rho_{\rm tot}$ is the error on the total density.} due to the poorer sampling.\\   
%To evaluate the entity of such discrepancy, we have computed the amount of stellar mass outside 1 kpc in the reference \citetalias{Pas18} model and for the end states of S1 and S2. We find that the semi-analytical models of the Fornax dSph has about $2.81\times 10^6M_\odot$ of stellar mass sitting at $r>1$ kpc, corresponding to a fraction of about the 30\% of the total stellar mass, while for S1 and S2 we computed a $5.71\times 10^6$ and $2.56\times 10^6$ solar masses, around the 32\% and 20\% respectively. 
%To test the stability of the given mass fraction portion function, we applied the same best-fit mass function for S1 and S2 to the isolated test models discussed in Sect. \ref{testsrho}. The resulting initial and final density profiles for DM and stars match in the worst case within a 5\% accuracy, corresponding to the inner region of the stellar density, i.e. the region with larger errors (cfr. also the error bars in the top panels of Fig. \ref{fig:S1-Density-Prof}) for the $N-$body realizations of both M1 and M2.\\
The tidal forces caused by the interaction with the MW gravitational field strip mass from the Fornax dSph (\citealt{2007MNRAS.381..987C,2015MNRAS.454.2401B}). To quantify the fractions of stellar and DM mass lost by the simulated satellite while orbiting the Galaxy, we evaluate as a function of time the respective mass content within two control spheres centred on the satellite's centre: in particular, as in \citet{2015MNRAS.454.2401B}, we choose spheres with radii $r_3=3$ kpc and $r_{1.6}=1.6$ kpc. In Fig. \ref{fig:S1-MassLoss} we present the time evolution of the mass fraction within $r_3$ and $r_{1.6}$ in simulations S1 and S2 for the stellar and DM components. 
Our results suggest that Fornax underwent significant DM mass loss, confirming the results  obtained by other authors (\citealt{2015MNRAS.454.2401B,Gen22}). 
However, in both simulations, only the DM mass is significantly affected by the tidal stripping, while the stellar mass remains essentially constant over 12 Gyr of evolution.
In simulation S1 there is a flux of stellar mass moving outward from the inner 1.6 kpc zone, but not leaving the sphere of radius $r_3$, the mass within which remains almost constant (see Fig.\ \ref{fig:S1-MassLoss}). 
However, the fraction of stellar mass lost in S1 within 1.6 kpc is only $\approx 4.8$\%; corresponding to roughly $5 \times 10^5 \text{ M}_{\odot}$. 
\begin{table}
\begin{tabular}{lllll}
 S1 & $M_\star$ [$\text{M}_{\odot}$] & \begin{tabular}[c]{@{}l@{}} \% \end{tabular} & $M_{DM}$ [$\text{M}_{\odot}$] & \begin{tabular}[c]{@{}l@{}} \%  \end{tabular}  \\ \hline
$M_{1.6 \text{ kpc}}(\text{t=0})$ & $1.43 \times 10^7$& & $5.57 \times 10^8$ &    \\
%$M_{1.6 \text{ kpc}}(\text{t=12})$ & $1.36 \times 10^7$ &(95\%) & $4.18 \times 10^8$ &(75\%)    \\
$ \Delta M_{1.6 \text{ kpc}}$ & $0.07 \times 10^7$ &(5\%) & $1.39 \times 10^8$ &(25\%)     \\
$M_{3.0 \text{ kpc}}(\text{t=0})$ & $1.51 \times 10^7$ && $2.07 \times 10^9$ &     \\
%$M_{3.0 \text{ kpc}}(\text{t=12})$ & $1.51 \times 10^7$ &($100\%$) & $1.62 \times 10^9$ &(78\%)    \\
$ \Delta M_{3.0 \text{ kpc}}$ & $0.001\times 10^7$ &($<1\%$) & $0.39 \times 10^9$ &(22\%)   \\ \hline
S2 & & &    \\ \hline
$M_{1.6 \text{ kpc}}(\text{t=0})$ & $1.30 \times 10^7$& & $8.93 \times 10^8$ &    \\
%$M_{1.6 \text{ kpc}}(\text{t=12})$ & $1.28 \times 10^7$ &(98.5\%) & $4.64 \times 10^8$ &(52\%)    \\
$ \Delta M_{1.6 \text{ kpc}}$ & $0.02 \times 10^7$ &(1.5\%) & $4.29 \times 10^8$ &(48\%)     \\
$M_{3.0 \text{ kpc}}(\text{t=0})$ & $1.31 \times 10^7$ && $3.30 \times 10^9$ &     \\
%$M_{3.0 \text{ kpc}}(\text{t=12})$ & $1.31 \times 10^7$ &($100\%$) & $1.80 \times 10^9$ &(55\%)    \\
$ \Delta M_{3.0 \text{ kpc}}$ & $0.001\times 10^7$ &($<1\%$) & $4.16 \times 10^9$ &(46\%)   \\ \hline
\end{tabular}
\caption{ Initial mass within 1.6 kpc ($M_{1.6 \text{ kpc}}$) and 3 kpc ($M_{3 \text{ kpc}}$), and corresponding mass loss $\Delta M\equiv M(t=12\,\text{Gyr})-M(t=0)$ for the stellar (second column) and DM (fourth column) components. 
The third and fifth columns indicate the corresponding per cent mass loss.}
\label{tab:S1-massloss}
\end{table}
In simulation S2 we observe a more evident loss of mass. DM is heavily stripped away, and at the end of the simulation only 52$\%$ of the DM mass is still bound to the galaxy (see Fig. \ref{fig:S1-MassLoss}). However, as in simulation S1, also in simulation S2 the
stellar component is hardly touched; only after 8.5 Gyr there is 
a non-negligible variation of the stars within 1.6 kpc, which however does
not lead to stellar mass loss beyond 3 kpc: the stellar mass within 3 kpc remains constant throughout the simulation. Remarkably, even though the MW model considered here is rather heavy and fixed across the 12 Gyr of simulation, the stellar mass loss remains negligible. The data on the mass loss are reported in Tab. \ref{tab:S1-massloss}.\\
\indent From the evolution of the enclosed mass function of DM it is evident that mass loss happens, as expected, mostly when the system approaches the pericentre (the corresponding times are indicated in Fig. \ref{fig:S1-MassLoss} by the vertical dotted lines). 
%Typically, when the simulated satellite passes closer to the centre of the Galactic potential well, tidal forces compress and stretch its mass distribution causing some particles to be excited to larger velocities and escape. 
The increase of the enclosed mass fraction in correspondence of the dotted lines, in particular for S2, might be related to the system encountering mass lost at the previous pericentre passages. 
%%%%%%%%%%%%%%%%%%%%%%%%%%%%%%%%%%%%%%%%%%%%%%%%%%%%%%%

\subsection{Evolution of the satellite central dark matter density $\rho_{150}$ for different pericentric radii}
\label{sec:rho150}

\begin{figure}
        \centering
        \includegraphics[width=\columnwidth]{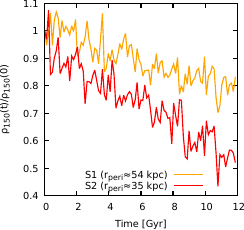}
\caption{Evolution of the satellite central DM density $\rho_{150}$, normalized to its value at $t=0$, in simulations S1 (yellow line) and S2 (red line).}
\label{fig150}
\end{figure}

In the classical dSphs satellites of the MW, the DM density at 150 pc from the centre of the dwarf galaxy 
(see e.g. \citealt{2019MNRAS.484.1401R}), $\rho_{150}$, is found to anticorrelate with the pericentric radius $r_{\rm peri}$  of the dwarf's orbit (\citealt{2019MNRAS.490..231K}). 
Though the robustness \citep{2023MNRAS.522.3058C} and interpretation \citep{And23} of this anticorrelation is debated, 
there is general agreement that the dynamical interaction between the dwarf satellites and the MW can contribute to determine the  relationship between DM density and orbital properties of the present-day dSphs. For instance, it has been suggested \citep[e.g.][]{2019MNRAS.490..231K} that an anticorrelation might be produced by the so-called survivor bias: among the dwarfs with small-$\rp$ orbits, only those with high central DM density manage to survive while the others are completely disrupted.\\
\indent  
Assuming the parameters of the portion function $\Pstar$ as given in Table  \ref{tab:SIM_PROP}, for both simulations S1 and S2 we measured $\rho_{150}$ and monitored its value as a function of time. 
The simulations presented in this work, in which the satellite always survives, cannot provide information on the survivor bias scenario. However, given that simulations S1 and S2 are characterized by significantly different $\rp$ (see Table \ref{tab:init}), but are designed to have similar final DM density distributions,  when interpreted in the context of the  $\rho_{150}$-$r_{\rm peri}$ anticorrelation, they can be used to illustrate, with a clean quantitative example, two important features to be considered when studying the origin the relationship between $\rho_{150}$ and $r_{\rm peri}$. 
\begin{enumerate}
\item Two MW dSphs with very similar present-day $\rho_{150}$ can have very different orbits (and values of $\rp$). In both simulation S1 (with $\rp\approx 54$ kpc) and simulation S2  (with $\rp\approx 35$ kpc) the final (i.e.\ present-day) central DM density is $\rho_{150}\approx 4\times 10^7M_\odot/{\rm pc}^3$ (see Fig.\ \ref{fig:S1-Density-Prof}). 
\item If the satellites are not completely disrupted, the tidal field of the MW might also have the effect of inducing a {\em correlation} between $\rho_{150}$ and $r_{\rm peri}$ \citep[see also ][]{2019MNRAS.490..231K}. This is illustrated by  Fig.\ \ref{fig150}, showing for both simulations S1 and S2, as a function of time, $\rho_{150}$ normalized to its initial value. In both cases $\rho_{150}$ decreases with time, but it is apparent that $\rho_{150}$ decreases at a faster rate in the simulation with lower $\rp$.  Given that the ratio of $\rp$ is $\approx1.6$ and the ratio of $\rho_{150}$ variation is $\approx 0.7$, this effect is opposite, but quantitatively similar, to the empirical anticorrelation \citep[see][]{2023MNRAS.522.3058C}.
\end{enumerate}
%%%%%%%%%%%%%%%%%%%%%%%%%
%%%%%%%%%%%%%%%%%%%%%%%%%%%%%%%%%%%%%%%
%%%%%%%%%%%%%%%%%%%%%%%%%%%%%%%%%%%%%%%%%%%%%%%%%%%%%%%%%%%%%%%%
\section{Discussion}
\label{sec:discussion} 
We have dedicated attention to the evolution of the total stellar mass of the Fornax dSph induced by the tidal field of the MW, 
which provides important constraints for scenarios for the origin of multiple populations (MPs) in GCs. We showed that, during its evolution, Fornax underwent very modest stellar mass loss, thus confirming the results of the simulations of \citet{2015MNRAS.454.2401B}.
However, it must be stressed that, for the purpose of constraining the stellar mass loss of Fornax, our result is more general than that of \citet{2015MNRAS.454.2401B}, who, in their simulations (set up in the standard way with a two component $N-$body model using stellar and DM particles) assumed an initial stellar density distribution of the satellite similar to that observed in Fornax. \\
\indent
In our effective two-component simulations, we fix the initial total density distribution of the satellite, but there is no a priori assumption on the stellar density distribution, apart from the relatively weak condition that $\rho_\star$ cannot exceed   $\rhotot$.  Thus, though solutions (i.e. choices of the function $\Pstar$) in which the initial stellar mass of Fornax was significantly higher than its present-day stellar mass would be possible outcomes of our simulations, it turns out that these solutions are excluded when we impose that the end-product of the simulation must have stellar density distribution consistent with the observations. 
We can thus robustly conclude that the Fornax dSph has retained most of its original stellar content.\\
\indent In the light of these findings, it remains open the question of explaining both 
the very high GC specific frequency and the possibility of accommodating the origin
of the MPs in GCs in the framework of self-enrichment scenarios, such as the AGB and fast rotating massive stars ones, which assume that, originally, GCs were much more massive than today. At this stage, it is important to stress that caution must be taken when analysing the properties of the GCs and field stars of Fornax. \\
\indent To estimate the amount of this component, \cite{2012A&A...544L..14L,2018A&A...613A..56L} assumed an average mass-to-light ratio $M/L_V = 3.5$ (\citealt{1992ApJ...400..510D}) to show that about 1/2 of the mass in metal-poor field stars is in GCs, and that the present-day GCs could at most have been by a factor $\sim 4-5$ more massive initially then they are now. These data are particularly impressive if compared to the ratio between the total mass in MW GCs and the stellar halo, of the order of a few per cent only. However, the total stellar mass 
of Fornax is still significantly affected by uncertainties, in particular regarding 
the assumed $M/L_V$ ratio, for which  SSP models for a standard IMF predict lower values, $M/L_V=2$ (e.g. \citealt{2011AJ....142....8S}).\\
\indent By means of up-to-date photometric data of Galactic GCs, 
the recent estimate of \cite{2020PASA...37...46B} indicates an average mass-to-light ratio of $M/L_V$=1.83. Still, this value is subject to various uncertainties, such as stellar age, metallicity  and the impact of internal dynamical processes, such as mass segregation (\citealt{2020PASA...37...46B} and references therein). 
However, if the properties of Galactic GCs were similar to the ones of their Fornax homologous, this $M/L_V$ value would imply a 1.9 increase in the ratio between field stars and GCs and an upper limit of 9.6 between the total GC mass and field stars, bringing it in better agreement with the values expected in the MPs formation scenarios discussed above.\\
\indent In the future, it will be important to develop 
new models to study the mass of such clusters throughout their history 
and, starting from a set of different initial conditions, which evolutionary path 
led to the current configuration. 
These models need to take into account the influence 
of the cosmological environment, to compute how the time-varying tidal field emerging from 
the past merging history has contributed 
to cause the dynamical evolution of the Fornax clusters, as already performed for MW GCs by \cite{2022ApJS..258...22R}. 
A significant result in this field is the recent one by \cite{2024MNRAS.527.2765M}, 
who showed how the disruption processes of GCs are sensitive to the mass and density of the host galaxy and supporting larger GC specific frequencies in dwarf galaxies. Moreover, future models should also include the effects of stellar evolution on the (baryonic) mass loss.  \\
\indent\citet{10.1093/mnras/stv2944} conjecture that few hyper-velocity stars escaping the GCs of Fornax might be accelerated by other evolutionary processes such as gas expulsion from GCs (the latter also contributing to the total mass loss). 
Indeed, other studies have indicated that intense stellar feedback may lead to significant loss of residual gas from GCs (e.g. \citealt{2007MNRAS.380.1589B,2015ApJ...814L..14C}, \citealt{2018MNRAS.478.5112S}). In the case of Fornax, to achieve the large velocities necessary for stellar expulsion, gas loss has to be significant and rapid, i.e. to occur on timescales $\le 10^5$ yr. 
Assessing this possibility requires dedicated simulations to explore a wide parameter space, including stellar feedback prescriptions, star formation efficiency and some GCs structural properties. 
Moreover, a further, necessary condition for the mechanism proposed by \citet{10.1093/mnras/stv2944} to be effective is that the Fornax dark halo is cuspy (or with a small core), which is not favoured by some recent dynamical models \citep[][and references therein]{Pas18,Bat22}. 
\section{Summary and conclusions}
\label{sec:conclusions}
%%%%%%%%%%%%%%%%%%%%%%%%%
We have studied the dynamical evolution of the Fornax dSph with effective multi-component $N-$body simulations. Using two different orbits in the MW potential, based on state-of-the art estimates of Fornax's orbital parameters, we find for both orbits considerable DM mass loss along 12 Gyr integration time, and rather mild orbital decay induced by dynamical self-friction.\\  
\indent At the end of the simulations, the simulated Fornax-like satellite is located in a position consistent with the present-day observed position of Fornax. The final stellar density profile of the simulated satellite is 
 (at least in one case) well within the error bars of the  model of \citetalias{Pas18}. The DM component suffers substantial depletion due to tidal stripping in the Galactic potential, losing a fraction of between the 20 and the 50\% within a 3 kpc radius.  The central DM density of the simulated satellite $\rho_{150}$, measured at 150 pc  from the centre, decreases more in the orbit with the smaller pericentric radius $\rp$. Thus, our simulations provide further indications that an $\rho_{150}$-$r_{\rm peri}$ anticorrelation is not a straightforward consequence of the dynamical interaction of the dwarf satellites with the tidal field of the MW.\\ 
 \indent The main result of this work is that, though our initial conditions would permit a wide variety of initial stellar mass distributions for Fornax, for the final stellar and DM density distributions to be consistent with the ones observed at the present day,    
 the stellar mass loss must have been very low ($\lesssim 3\%$). Thus, it appears rather unlikely that loss of mass from Fornax to the MW could be the explanation of the anomalously high fraction of stellar mass that in Fornax resides in GCs, which remains a problem (despite the significant uncertainty in the adopted GC mass-to-light ratio) 
 for GC formation scenarios that postulate that GCs were originally much more massive. 
%%%%%%%%%%%%%%%%%%%%%%%%%%%%%%%%%%%%%%%%%%%%%%%%%%
\begin{acknowledgements}
PFDC wishes to acknowledge funding by ‘‘Fondazione Cassa di Risparmio di Firenze" under the project {\it HIPERCRHEL} for the use of high performance computing resources at the University of Firenze. The research activities described in this paper have been co-funded by the European Union – NextGenerationEU within PRIN 2022 project n.20229YBSAN - Globular clusters in cosmological simulations and in lensed fields: from their birth to the present epoch.
G.I. acknowledges financial support under the National Recovery and Resilience Plan (NRRP), Mission 4, Component 2, Investment 1.4, - Call for tender No. 3138 of 18/12/2021 of Italian Ministry of
University and Research funded by the European Union – NextGenerationEU.
\end{acknowledgements}
   \bibliographystyle{aa} 
   \bibliography{biblio_fnxgc} 

\begin{thebibliography}{82}
\expandafter\ifx\csname natexlab\endcsname\relax\def\natexlab#1{#1}\fi

\bibitem[{{Andrade} {et~al.}(2023){Andrade}, {Kaplinghat}, \& {Valli}}]{And23}
{Andrade}, K.~E., {Kaplinghat}, M., \& {Valli}, M. 2023, arXiv e-prints,
  arXiv:2311.01528

\bibitem[{{Battaglia} {et~al.}(2005){Battaglia}, {Helmi}, {Morrison},
  {Harding}, {Olszewski}, {Mateo}, {Freeman}, {Norris}, \&
  {Shectman}}]{2005MNRAS.364..433B}
{Battaglia}, G., {Helmi}, A., {Morrison}, H., {et~al.} 2005, \mnras, 364, 433

\bibitem[{{Battaglia} \& {Nipoti}(2022)}]{Bat22}
{Battaglia}, G. \& {Nipoti}, C. 2022, Nature Astronomy, 6, 659

\bibitem[{{Battaglia} {et~al.}(2015){Battaglia}, {Sollima}, \&
  {Nipoti}}]{2015MNRAS.454.2401B}
{Battaglia}, G., {Sollima}, A., \& {Nipoti}, C. 2015, \mnras, 454, 2401

\bibitem[{{Battaglia} {et~al.}(2022){Battaglia}, {Taibi}, {Thomas}, \&
  {Fritz}}]{GAIADR3}
{Battaglia}, G., {Taibi}, S., {Thomas}, G.~F., \& {Fritz}, T.~K. 2022, \aap,
  657, A54

\bibitem[{{Battaglia} {et~al.}(2006){Battaglia}, {Tolstoy}, {Helmi}, {Irwin},
  {Letarte}, {Jablonka}, {Hill}, {Venn}, {Shetrone}, {Arimoto}, {Primas},
  {Kaufer}, {Francois}, {Szeifert}, {Abel}, \& {Sadakane}}]{Bat06}
{Battaglia}, G., {Tolstoy}, E., {Helmi}, A., {et~al.} 2006, \aap, 459, 423

\bibitem[{{Baumgardt} \& {Kroupa}(2007)}]{2007MNRAS.380.1589B}
{Baumgardt}, H. \& {Kroupa}, P. 2007, \mnras, 380, 1589

\bibitem[{{Baumgardt} {et~al.}(2020){Baumgardt}, {Sollima}, \&
  {Hilker}}]{2020PASA...37...46B}
{Baumgardt}, H., {Sollima}, A., \& {Hilker}, M. 2020, \pasa, 37, e046

\bibitem[{{Binney}(1981)}]{1981MNRAS.196..455B}
{Binney}, J. 1981, \mnras, 196, 455

\bibitem[{{Bird} {et~al.}(2021){Bird}, {Xue}, {Liu}, {Shen}, {Flynn}, {Yang},
  {Zhao}, \& {Tian}}]{2021ApJ...919...66B}
{Bird}, S.~A., {Xue}, X.-X., {Liu}, C., {et~al.} 2021, \apj, 919, 66

\bibitem[{{Borukhovetskaya} {et~al.}(2022){Borukhovetskaya}, {Errani},
  {Navarro}, {Fattahi}, \& {Santos-Santos}}]{Bor22}
{Borukhovetskaya}, A., {Errani}, R., {Navarro}, J.~F., {Fattahi}, A., \&
  {Santos-Santos}, I. 2022, \mnras, 509, 5330

\bibitem[{{Bovy}(2015)}]{Bov15}
{Bovy}, J. 2015, \apjs, 216, 29

\bibitem[{{Bullock} \& {Johnston}(2005)}]{Bul05}
{Bullock}, J.~S. \& {Johnston}, K.~V. 2005, \apj, 635, 931

\bibitem[{{Calura} {et~al.}(2014){Calura}, {Ciotti}, \&
  {Nipoti}}]{2014MNRAS.440.3341C}
{Calura}, F., {Ciotti}, L., \& {Nipoti}, C. 2014, \mnras, 440, 3341

\bibitem[{{Calura} {et~al.}(2019){Calura}, {D'Ercole}, {Vesperini}, {Vanzella},
  \& {Sollima}}]{2019MNRAS.489.3269C}
{Calura}, F., {D'Ercole}, A., {Vesperini}, E., {Vanzella}, E., \& {Sollima}, A.
  2019, \mnras, 489, 3269

\bibitem[{{Calura} {et~al.}(2015){Calura}, {Few}, {Romano}, \&
  {D'Ercole}}]{2015ApJ...814L..14C}
{Calura}, F., {Few}, C.~G., {Romano}, D., \& {D'Ercole}, A. 2015, \apjl, 814,
  L14

\bibitem[{{Cardona-Barrero} {et~al.}(2023){Cardona-Barrero}, {Battaglia},
  {Nipoti}, \& {Di Cintio}}]{2023MNRAS.522.3058C}
{Cardona-Barrero}, S., {Battaglia}, G., {Nipoti}, C., \& {Di Cintio}, A. 2023,
  \mnras, 522, 3058

\bibitem[{{Choi} {et~al.}(2007){Choi}, {Weinberg}, \&
  {Katz}}]{2007MNRAS.381..987C}
{Choi}, J.-H., {Weinberg}, M.~D., \& {Katz}, N. 2007, \mnras, 381, 987

\bibitem[{{Contigiani} {et~al.}(2019){Contigiani}, {Rossi}, \&
  {Marchetti}}]{2019MNRAS.487.4025C}
{Contigiani}, O., {Rossi}, E.~M., \& {Marchetti}, T. 2019, \mnras, 487, 4025

\bibitem[{{D'Antona} {et~al.}(2016){D'Antona}, {Vesperini}, {D'Ercole},
  {Ventura}, {Milone}, {Marino}, \& {Tailo}}]{2016MNRAS.458.2122D}
{D'Antona}, F., {Vesperini}, E., {D'Ercole}, A., {et~al.} 2016, \mnras, 458,
  2122

\bibitem[{{Darragh-Ford} {et~al.}(2021){Darragh-Ford}, {Nadler}, {McLaughlin},
  \& {Wechsler}}]{2021ApJ...915...48D}
{Darragh-Ford}, E., {Nadler}, E.~O., {McLaughlin}, S., \& {Wechsler}, R.~H.
  2021, \apj, 915, 48

\bibitem[{{de Boer} {et~al.}(2012){de Boer}, {Tolstoy}, {Hill}, {Saha},
  {Olszewski}, {Mateo}, {Starkenburg}, {Battaglia}, \& {Walker}}]{deB12}
{de Boer}, T.~J.~L., {Tolstoy}, E., {Hill}, V., {et~al.} 2012, \aap, 544, A73

\bibitem[{{Decressin} {et~al.}(2008){Decressin}, {Baumgardt}, \&
  {Kroupa}}]{2008A&A...492..101D}
{Decressin}, T., {Baumgardt}, H., \& {Kroupa}, P. 2008, \aap, 492, 101

\bibitem[{{Decressin} {et~al.}(2007){Decressin}, {Meynet}, {Charbonnel},
  {Prantzos}, \& {Ekstr{\"o}m}}]{2007A&A...464.1029D}
{Decressin}, T., {Meynet}, G., {Charbonnel}, C., {Prantzos}, N., \&
  {Ekstr{\"o}m}, S. 2007, \aap, 464, 1029

\bibitem[{{Dehnen}(2002)}]{2002JCoPh.179...27D}
{Dehnen}, W. 2002, Journal of Computational Physics, 179, 27

\bibitem[{{D'Ercole} {et~al.}(2008){D'Ercole}, {Vesperini}, {D'Antona},
  {McMillan}, \& {Recchi}}]{2008MNRAS.391..825D}
{D'Ercole}, A., {Vesperini}, E., {D'Antona}, F., {McMillan}, S. L.~W., \&
  {Recchi}, S. 2008, \mnras, 391, 825

\bibitem[{{Dubath} {et~al.}(1992){Dubath}, {Meylan}, \&
  {Mayor}}]{1992ApJ...400..510D}
{Dubath}, P., {Meylan}, G., \& {Mayor}, M. 1992, \apj, 400, 510

\bibitem[{{Elias} {et~al.}(2020){Elias}, {Sales}, {Helmi}, \&
  {Hernquist}}]{2020MNRAS.495...29E}
{Elias}, L.~M., {Sales}, L.~V., {Helmi}, A., \& {Hernquist}, L. 2020, \mnras,
  495, 29

\bibitem[{{Elmegreen}(1999)}]{1999Ap&SS.269..469E}
{Elmegreen}, B.~G. 1999, \apss, 269, 469

\bibitem[{{Errani} {et~al.}(2015){Errani}, {Penarrubia}, \& {Tormen}}]{Err15}
{Errani}, R., {Penarrubia}, J., \& {Tormen}, G. 2015, \mnras, 449, L46

\bibitem[{Garavito-Camargo {et~al.}(2021)Garavito-Camargo, Besla, Laporte,
  Price-Whelan, Cunningham, Johnston, Weinberg, \&
  Gómez}]{Garavito-Camargo_2021}
Garavito-Camargo, N., Besla, G., Laporte, C. F.~P., {et~al.} 2021, The
  Astrophysical Journal, 919, 109

\bibitem[{{Genina} {et~al.}(2022){Genina}, {Read}, {Fattahi}, \&
  {Frenk}}]{Gen22}
{Genina}, A., {Read}, J.~I., {Fattahi}, A., \& {Frenk}, C.~S. 2022, \mnras,
  510, 2186

\bibitem[{{Gibbons} {et~al.}(2014){Gibbons}, {Belokurov}, \&
  {Evans}}]{2014MNRAS.445.3788G}
{Gibbons}, S.~L.~J., {Belokurov}, V., \& {Evans}, N.~W. 2014, \mnras, 445, 3788

\bibitem[{{GRAVITY Collaboration} {et~al.}(2018){GRAVITY Collaboration},
  {Abuter}, {Amorim}, {Anugu}, {Baub{\"o}ck}, {Benisty}, {Berger}, {Blind},
  {Bonnet}, {Brandner}, {Buron}, {Collin}, {Chapron}, {Cl{\'e}net}, {Coud{\'e}
  Du Foresto}, {de Zeeuw}, {Deen}, {Delplancke-Str{\"o}bele}, {Dembet},
  {Dexter}, {Duvert}, {Eckart}, {Eisenhauer}, {Finger}, {F{\"o}rster
  Schreiber}, {F{\'e}dou}, {Garcia}, {Garcia Lopez}, {Gao}, {Gendron},
  {Genzel}, {Gillessen}, {Gordo}, {Habibi}, {Haubois}, {Haug}, {Hau{\ss}mann},
  {Henning}, {Hippler}, {Horrobin}, {Hubert}, {Hubin}, {Jimenez Rosales},
  {Jochum}, {Jocou}, {Kaufer}, {Kellner}, {Kendrew}, {Kervella}, {Kok},
  {Kulas}, {Lacour}, {Lapeyr{\`e}re}, {Lazareff}, {Le Bouquin}, {L{\'e}na},
  {Lippa}, {Lenzen}, {M{\'e}rand}, {M{\"u}ler}, {Neumann}, {Ott}, {Palanca},
  {Paumard}, {Pasquini}, {Perraut}, {Perrin}, {Pfuhl}, {Plewa}, {Rabien},
  {Ram{\'\i}rez}, {Ramos}, {Rau}, {Rodr{\'\i}guez-Coira}, {Rohloff}, {Rousset},
  {Sanchez-Bermudez}, {Scheithauer}, {Sch{\"o}ller}, {Schuler}, {Spyromilio},
  {Straub}, {Straubmeier}, {Sturm}, {Tacconi}, {Tristram}, {Vincent}, {von
  Fellenberg}, {Wank}, {Waisberg}, {Widmann}, {Wieprecht}, {Wiest},
  {Wiezorrek}, {Woillez}, {Yazici}, {Ziegler}, \& {Zins}}]{gravity2018}
{GRAVITY Collaboration}, {Abuter}, R., {Amorim}, A., {et~al.} 2018, \aap, 615,
  L15

\bibitem[{{Hernquist}(1990)}]{Her90}
{Hernquist}, L. 1990, \apj, 356, 359

\bibitem[{{Huang} \& {Koposov}(2021)}]{2021MNRAS.500..986H}
{Huang}, K.-W. \& {Koposov}, S.~E. 2021, \mnras, 500, 986

\bibitem[{{Iorio} {et~al.}(2019){Iorio}, {Nipoti}, {Battaglia}, \&
  {Sollima}}]{Ior19}
{Iorio}, G., {Nipoti}, C., {Battaglia}, G., \& {Sollima}, A. 2019, \mnras, 487,
  5692

\bibitem[{{Johnston} {et~al.}(1995){Johnston}, {Spergel}, \&
  {Hernquist}}]{Joh95}
{Johnston}, K.~V., {Spergel}, D.~N., \& {Hernquist}, L. 1995, \apj, 451, 598

\bibitem[{{Kaplinghat} {et~al.}(2019){Kaplinghat}, {Valli}, \&
  {Yu}}]{2019MNRAS.490..231K}
{Kaplinghat}, M., {Valli}, M., \& {Yu}, H.-B. 2019, \mnras, 490, 231

\bibitem[{{Kazantzidis} {et~al.}(2004){Kazantzidis}, {Magorrian}, \&
  {Moore}}]{Kaz04}
{Kazantzidis}, S., {Magorrian}, J., \& {Moore}, B. 2004, \apj, 601, 37

\bibitem[{Khalaj \& Baumgardt(2016)}]{10.1093/mnras/stv2944}
Khalaj, P. \& Baumgardt, H. 2016, Monthly Notices of the Royal Astronomical
  Society, 457, 479

\bibitem[{{Koposov} {et~al.}(2023){Koposov}, {Erkal}, {Li}, {Da Costa},
  {Cullinane}, {Ji}, {Kuehn}, {Lewis}, {Pace}, {Shipp}, {Zucker},
  {Bland-Hawthorn}, {Lilleengen}, {Martell}, \& {S5
  Collaboration}}]{2023MNRAS.521.4936K}
{Koposov}, S.~E., {Erkal}, D., {Li}, T.~S., {et~al.} 2023, \mnras, 521, 4936

\bibitem[{{Kowalczyk} {et~al.}(2019){Kowalczyk}, {del Pino}, {{\L}okas}, \&
  {Valluri}}]{Kow19}
{Kowalczyk}, K., {del Pino}, A., {{\L}okas}, E.~L., \& {Valluri}, M. 2019,
  \mnras, 482, 5241

\bibitem[{{Kroupa}(2001)}]{2001MNRAS.322..231K}
{Kroupa}, P. 2001, \mnras, 322, 231

\bibitem[{{Kruijssen} {et~al.}(2019){Kruijssen}, {Pfeffer}, {Reina-Campos},
  {Crain}, \& {Bastian}}]{2019MNRAS.486.3180K}
{Kruijssen}, J.~M.~D., {Pfeffer}, J.~L., {Reina-Campos}, M., {Crain}, R.~A., \&
  {Bastian}, N. 2019, \mnras, 486, 3180

\bibitem[{{Larsen} {et~al.}(2018){Larsen}, {Brodie}, {Wasserman}, \&
  {Strader}}]{2018A&A...613A..56L}
{Larsen}, S.~S., {Brodie}, J.~P., {Wasserman}, A., \& {Strader}, J. 2018, \aap,
  613, A56

\bibitem[{{Larsen} {et~al.}(2012){Larsen}, {Strader}, \&
  {Brodie}}]{2012A&A...544L..14L}
{Larsen}, S.~S., {Strader}, J., \& {Brodie}, J.~P. 2012, \aap, 544, L14

\bibitem[{{Letarte} {et~al.}(2006){Letarte}, {Hill}, {Jablonka}, {Tolstoy},
  {Fran{\c{c}}ois}, \& {Meylan}}]{2006A&A...453..547L}
{Letarte}, B., {Hill}, V., {Jablonka}, P., {et~al.} 2006, \aap, 453, 547

\bibitem[{{Lilleengen} {et~al.}(2023){Lilleengen}, {Petersen}, {Erkal},
  {Pe{\~n}arrubia}, {Koposov}, {Li}, {Cullinane}, {Ji}, {Kuehn}, {Lewis},
  {Mackey}, {Pace}, {Shipp}, {Zucker}, {Bland-Hawthorn}, {Hilmi}, \& {S5
  Collaboration}}]{2023MNRAS.518..774L}
{Lilleengen}, S., {Petersen}, M.~S., {Erkal}, D., {et~al.} 2023, \mnras, 518,
  774

\bibitem[{{Londrillo} {et~al.}(2003){Londrillo}, {Nipoti}, \&
  {Ciotti}}]{2003MSAIS...1...18L}
{Londrillo}, P., {Nipoti}, C., \& {Ciotti}, L. 2003, Memorie della Societa
  Astronomica Italiana Supplementi, 1, 18

\bibitem[{{Massari} {et~al.}(2023){Massari}, {Aguado-Agelet}, {Monelli},
  {Cassisi}, {Pancino}, {Saracino}, {Gallart}, {Ruiz-Lara},
  {Fern{\'a}ndez-Alvar}, {Surot}, {Stokholm}, {Salaris}, {Miglio}, \&
  {Ceccarelli}}]{2023A&A...680A..20M}
{Massari}, D., {Aguado-Agelet}, F., {Monelli}, M., {et~al.} 2023, \aap, 680,
  A20

\bibitem[{{Miller} {et~al.}(2020){Miller}, {van den Bosch}, {Green}, \&
  {Ogiya}}]{Mil20}
{Miller}, T.~B., {van den Bosch}, F.~C., {Green}, S.~B., \& {Ogiya}, G. 2020,
  \mnras, 495, 4496

\bibitem[{{Milone} {et~al.}(2015){Milone}, {Marino}, {Piotto}, {Bedin},
  {Anderson}, {Renzini}, {King}, {Bellini}, {Brown}, {Cassisi}, {D'Antona},
  {Jerjen}, {Nardiello}, {Salaris}, {Marel}, {Vesperini}, {Yong}, {Aparicio},
  {Sarajedini}, \& {Zoccali}}]{2015MNRAS.447..927M}
{Milone}, A.~P., {Marino}, A.~F., {Piotto}, G., {et~al.} 2015, \mnras, 447, 927

\bibitem[{{Miyamoto} \& {Nagai}(1975)}]{MiyamotoNagai}
{Miyamoto}, M. \& {Nagai}, R. 1975, \pasj, 27, 533

\bibitem[{{Moreno-Hilario} {et~al.}(2024){Moreno-Hilario}, {Martinez-Medina},
  {Li}, {Souza}, \& {P{\'e}rez-Villegas}}]{2024MNRAS.527.2765M}
{Moreno-Hilario}, E., {Martinez-Medina}, L.~A., {Li}, H., {Souza}, S.~O., \&
  {P{\'e}rez-Villegas}, A. 2024, \mnras, 527, 2765

\bibitem[{{Mori} {et~al.}(2024){Mori}, {Di Matteo}, {Salvadori}, {Khoperskov},
  {Pagnini}, \& {Haywood}}]{2024arXiv240113737M}
{Mori}, A., {Di Matteo}, P., {Salvadori}, S., {et~al.} 2024, arXiv e-prints,
  arXiv:2401.13737

\bibitem[{{Mucciarelli} \& {Massari}(2023)}]{2023MmSAI..94b.173M}
{Mucciarelli}, A. \& {Massari}, D. 2023, in Memorie della Societa Astronomica
  Italiana, Vol.~94, 173

\bibitem[{{Mulder}(1983)}]{Mulder}
{Mulder}, W.~A. 1983, aap, 117, 9

\bibitem[{{Muni} {et~al.}(2024){Muni}, {Pontzen}, {Read}, {Agertz}, {Rey}, \&
  {Taylor}}]{Mun24}
{Muni}, C., {Pontzen}, A., {Read}, J.~I., {et~al.} 2024, arXiv e-prints,
  arXiv:2407.14579

\bibitem[{{Nipoti} \& {Binney}(2015)}]{Nip15}
{Nipoti}, C. \& {Binney}, J. 2015, \mnras, 446, 1820

\bibitem[{{Nipoti} {et~al.}(2021){Nipoti}, {Cherchi}, {Iorio}, \&
  {Calura}}]{Nip21}
{Nipoti}, C., {Cherchi}, G., {Iorio}, G., \& {Calura}, F. 2021, \mnras, 503,
  4221

\bibitem[{{Nipoti} {et~al.}(2003){Nipoti}, {Londrillo}, \&
  {Ciotti}}]{2003MNRAS.342..501N}
{Nipoti}, C., {Londrillo}, P., \& {Ciotti}, L. 2003, \mnras, 342, 501

\bibitem[{{Nipoti} {et~al.}(2006){Nipoti}, {Londrillo}, \&
  {Ciotti}}]{2006MNRAS.370..681N}
{Nipoti}, C., {Londrillo}, P., \& {Ciotti}, L. 2006, \mnras, 370, 681

\bibitem[{{Pace} {et~al.}(2021){Pace}, {Walker}, {Koposov}, {Caldwell},
  {Mateo}, {Olszewski}, {Bailey}, \& {Wang}}]{2021ApJ...923...77P}
{Pace}, A.~B., {Walker}, M.~G., {Koposov}, S.~E., {et~al.} 2021, \apj, 923, 77

\bibitem[{{Pascale} {et~al.}(2018){Pascale}, {Posti}, {Nipoti}, \&
  {Binney}}]{Pas18}
{Pascale}, R., {Posti}, L., {Nipoti}, C., \& {Binney}, J. 2018, \mnras, 480,
  927

\bibitem[{{Petts} {et~al.}(2016){Petts}, {Read}, \&
  {Gualandris}}]{2016MNRAS.463..858P}
{Petts}, J.~A., {Read}, J.~I., \& {Gualandris}, A. 2016, \mnras, 463, 858

\bibitem[{{Piffl} {et~al.}(2014){Piffl}, {Binney}, {McMillan}, {Steinmetz},
  {Helmi}, {Wyse}, {Bienaym{\'e}}, {Bland-Hawthorn}, {Freeman}, {Gibson},
  {Gilmore}, {Grebel}, {Kordopatis}, {Navarro}, {Parker}, {Reid}, {Seabroke},
  {Siebert}, {Watson}, \& {Zwitter}}]{2014MNRAS.445.3133P}
{Piffl}, T., {Binney}, J., {McMillan}, P.~J., {et~al.} 2014, \mnras, 445, 3133

\bibitem[{{Power} {et~al.}(2003){Power}, {Navarro}, {Jenkins}, {Frenk},
  {White}, {Springel}, {Stadel}, \& {Quinn}}]{Pow03}
{Power}, C., {Navarro}, J.~F., {Jenkins}, A., {et~al.} 2003, \mnras, 338, 14

\bibitem[{{Read} {et~al.}(2019){Read}, {Walker}, \&
  {Steger}}]{2019MNRAS.484.1401R}
{Read}, J.~I., {Walker}, M.~G., \& {Steger}, P. 2019, \mnras, 484, 1401

\bibitem[{{Read} {et~al.}(2006){Read}, {Wilkinson}, {Evans}, {Gilmore}, \&
  {Kleyna}}]{Rea06}
{Read}, J.~I., {Wilkinson}, M.~I., {Evans}, N.~W., {Gilmore}, G., \& {Kleyna},
  J.~T. 2006, \mnras, 366, 429

\bibitem[{{Rodriguez} {et~al.}(2022){Rodriguez}, {Weatherford}, {Coughlin},
  {Amaro-Seoane}, {Breivik}, {Chatterjee}, {Fragione}, {K{\i}ro{\u{g}}lu},
  {Kremer}, {Rui}, {Ye}, {Zevin}, \& {Rasio}}]{2022ApJS..258...22R}
{Rodriguez}, C.~L., {Weatherford}, N.~C., {Coughlin}, S.~C., {et~al.} 2022,
  \apjs, 258, 22

\bibitem[{{Salpeter}(1955)}]{1955ApJ...121..161S}
{Salpeter}, E.~E. 1955, \apj, 121, 161

\bibitem[{{Schechter}(1976)}]{1976ApJ...203..297S}
{Schechter}, P. 1976, \apj, 203, 297

\bibitem[{{Sch{\"o}nrich}(2012)}]{Schonrich2012}
{Sch{\"o}nrich}, R. 2012, \mnras, 427, 274

\bibitem[{{Sch{\"o}nrich} {et~al.}(2010){Sch{\"o}nrich}, {Binney}, \&
  {Dehnen}}]{Schonrich2010}
{Sch{\"o}nrich}, R., {Binney}, J., \& {Dehnen}, W. 2010, \mnras, 403, 1829

\bibitem[{{Shchelkanova} {et~al.}(2021){Shchelkanova}, {Hayashi}, \&
  {Blinnikov}}]{2021ApJ...909..147S}
{Shchelkanova}, G., {Hayashi}, K., \& {Blinnikov}, S. 2021, \apj, 909, 147

\bibitem[{{Silich} \& {Tenorio-Tagle}(2018)}]{2018MNRAS.478.5112S}
{Silich}, S. \& {Tenorio-Tagle}, G. 2018, \mnras, 478, 5112

\bibitem[{{Strader} {et~al.}(2011){Strader}, {Caldwell}, \&
  {Seth}}]{2011AJ....142....8S}
{Strader}, J., {Caldwell}, N., \& {Seth}, A.~C. 2011, \aj, 142, 8

\bibitem[{{Vasiliev}(2024)}]{2024MNRAS.527..437V}
{Vasiliev}, E. 2024, \mnras, 527, 437

\bibitem[{{Vasiliev} {et~al.}(2021){Vasiliev}, {Belokurov}, \&
  {Erkal}}]{2021MNRAS.501.2279V}
{Vasiliev}, E., {Belokurov}, V., \& {Erkal}, D. 2021, \mnras, 501, 2279

\bibitem[{{Wang} {et~al.}(2019){Wang}, {Koposov}, {Drlica-Wagner}, {Pieres},
  {Li}, {de Boer}, {Bechtol}, {Belokurov}, {Pace}, {Bacon}, {Abbott}, {Annis},
  {Bertin}, {Brooks}, {Buckley-Geer}, {Burke}, {Carnero Rosell}, {Carrasco
  Kind}, {Carretero}, {da Costa}, {De Vicente}, {Desai}, {Diehl}, {Doel},
  {Estrada}, {Flaugher}, {Fosalba}, {Frieman}, {Garc{\'\i}a-Bellido}, {Gerdes},
  {Gruen}, {Gruendl}, {Gschwend}, {Gutierrez}, {Hollowood}, {Honscheid},
  {Hoyle}, {James}, {Kent}, {Kuehn}, {Kuropatkin}, {Maia}, {Marshall},
  {Menanteau}, {Miquel}, {Plazas}, {Sanchez}, {Santiago}, {Scarpine},
  {Schindler}, {Schubnell}, {Serrano}, {Sevilla-Noarbe}, {Smith}, {Smith},
  {Sobreira}, {Suchyta}, {Swanson}, {Tarle}, {Thomas}, {Tucker}, {Walker}, \&
  {DES Collaboration}}]{2019ApJ...875L..13W}
{Wang}, M.~Y., {Koposov}, S., {Drlica-Wagner}, A., {et~al.} 2019, \apjl, 875,
  L13

\bibitem[{{Wang} {et~al.}(2016){Wang}, {Strigari}, {Lovell}, {Frenk}, \&
  {Zentner}}]{2016MNRAS.457.4248W}
{Wang}, M.-Y., {Strigari}, L.~E., {Lovell}, M.~R., {Frenk}, C.~S., \&
  {Zentner}, A.~R. 2016, \mnras, 457, 4248

\end{thebibliography}
\begin{appendix}
\section{Dynamical friction tests}
\label{sec:app_dyn_fric}
%%%%%%%%%%%%%%%%%%%%%%%%%%%%%%%%%%%%%%%
\begin{figure*}
        \centering
        \includegraphics[width=0.85\textwidth]{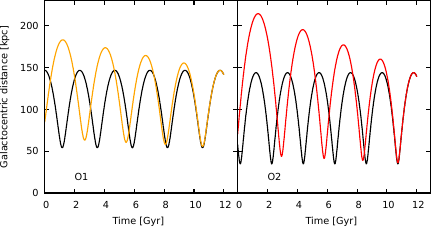}
\caption{ Left panel. Galactocentric distance as a function of time for the orbit O1 (black curve) and for a particle of mass $M_{\rm tot}= 7.2\times 10^9\Msun$ orbiting in the same gravitational potential and having the same phase-space coordinates at 12 Gyr, but under the effect of dynamical friction (yellow curve). Right panel: same as left panel, but for the orbit O2 (black curve) and the corresponding orbit with dynamical friction (red curve) for a particle of mass $M_{\rm tot}=1.13\times 10^{10}\Msun$.}
\label{fig:df}
\end{figure*}
%%%%%%%%%%%%%%%%%%%%%%%%%%%%%%%%%%%%%%%%%%%%%%%%%%%%%%%%%%%%%%%%
To test the possible influence of dynamical friction on the orbit of the Fornax dSph, we performed additional single particle integrations using {\sc galpy}, where a test mass $M_{\rm tot}$ orbits according to the equations of motion
\begin{equation}\label{eomdynfric}
\ddot{\mathbf{r}}=-\nabla\Phi_{\rm MW,tot}(\mathbf{r})+\mathbf{a}_{DF},
\end{equation}
where 
\begin{equation}\label{dynfric}
\mathbf{a}_{DF}=-2\pi G^2 M_{\rm tot}\rho(\mathbf{r})\ln(1+\Lambda^2)\Bigg[{\rm erf}(X)-\frac{2X^2}{\sqrt{\pi}}\exp(-X^2)\Bigg]\frac{\mathbf{v}}{v^3}
\end{equation}
is the deceleration produced by dynamical friction (\citealt{2016MNRAS.463..858P}).
Here $\rho(\mathbf{r})$ is the total mass density of the MW at $\mathbf{r}$, $\mathbf{v}=\dot{\mathbf{r}}$ and $X=v/(\sqrt{2}\sigma)$, with $\sigma(\mathbf{r})$ the velocity dispersion at ${\mathbf{r}}$ (i.e. the local Maxwellian approximation is assumed) evaluated solving the Jeans equations for the MW potential and density (see \citealt{Bov15} and references therein).
In Equation~(\ref{dynfric}), in order to take into account that the satellite is extended\footnote{ A more refined way to evaluate the dynamical friction on an extended object is given by the model of \cite{Mulder}, in which the satellite suffers a drag force that is augmented by the interaction between its density profile and the trailing wake. The satellite deformation and possibly the mass stripping induced by the background density of the parent galaxy are also accounted for, leading to a contribution to the slowing down that is however considerably lower than that of the pure dynamical friction.}, $\Lambda$ is defined as
\begin{equation}
\Lambda=\frac{r_{\rm Gal}}{{\rm max}\left [r_{\rm sat},\frac{G M_{\rm tot}}{v^2}\right]},
\end{equation}
where $r_{\rm Gal}$ and $r_{\rm sat}$ are the scale radii of the Galaxy and the satellite, which we take to be 12 kpc and 4.3 kpc, respectively.\\
\indent In the left panel of Figure \ref{fig:df}  we show the
Galactocentric distance as a function of time for the orbit O1  and for a particle of mass $M_{\rm tot}= 7.2\times 10^9\Msun$ (the total mass of model M1; Section~\ref{sec:nbody_models}) orbiting in the same gravitational potential and having the same phase-space coordinates at 12 Gyr, but under the effect of dynamical friction (Eq.\ \ref{eomdynfric}).  The analogous comparison for the orbit O2 is shown in the right panel of Figure \ref{fig:df}, where the orbit with dynamical friction is for a particle of mass $M_{\rm tot}=1.13\times 10^{10}\Msun$ (the total mass of model M2; Section~\ref{sec:nbody_models}). The Galactocentric distances for both models match rather well the parent trajectories without dynamical friction for, at least, the last 3 Gyr. 
At earlier times the orbits with and without dynamical friction differ significantly: the initial apocentre is larger in the presence than in the absence of dynamical friction, of about a factor 1.25 in the case of O1 and a factor 1.5 in the case of O2.\\
\indent As expected \citep{Mil20}, in the single particle orbital integration including dynamical friction, the orbital decay of the Fornax dSph is significantly larger than that in the $N-$body experiments including only dynamical self-friction (see Fig.\ \ref{fig:S1-CmVStime}). However, the pericentre, where most of the mass stripping happens, is almost unaffected by dynamical friction. We should also note that, since in our experiments the point mass $M_{\rm tot}$, representing the total mass of the satellite, is kept constant, the strength of the dynamical friction is systematically overestimated. To explore this processes more in detail one should run full $N-$body simulations where the MW is also modelled using particles, in the same spirit as, for example, \cite{2024arXiv240113737M}.
\end{appendix}
\end{document}